\documentclass[submission, Phys]{SciPost}
\usepackage{graphicx}
\usepackage{amssymb}
\usepackage{amsmath}
\usepackage{epsfig}
\usepackage{wasysym}
\usepackage{ulem}
\usepackage{color}
\usepackage{hyperref}
\usepackage{longtable}
\usepackage{bm}
\begin{document}
\begin{center}{\Large \textbf{
Invariants of winding-numbers and steric obstruction in dynamics of flux lines
}}\end{center}

\begin{center}
O. C\'epas\textsuperscript{1*},
P. M. Akhmetiev\textsuperscript{2,3}
\end{center}

\begin{center}
{\bf 1} Universit\'e Grenoble Alpes, CNRS, Grenoble INP, Institut N\'eel, 38000 Grenoble, France 
\\
{\bf 2}  HSE Tikhonov Moscow Institute of Electronics and Mathematics, 34 Tallinskaya Str., 123458, Moscow,  Russia 
\\
{\bf 3} Pushkov Institute of Terrestrial Magnetism, Ionosphere and Radio Wave Propagation, Kaluzhskoe Hwy 4, 108840, Moscow, Troitsk, Russia
\\
* olivier.cepas@neel.cnrs.fr
\end{center}

\begin{center}
\today
\end{center}


\section*{Abstract}
{\bf
  We classify the sectors of configurations that result from the dynamics of 2d crossing flux lines, which are the simplest degrees of freedom of the 3-coloring lattice model. We show that the dynamical obstruction is the consequence of two effects: (i) conservation laws described by a set of invariants that are polynomials of the winding numbers of the loop configuration, (ii) steric obstruction that prevents paths between configurations, for lack of free space.
  We argue that the invariants fully classify the configurations in five, chiral and achiral, sectors and no further obstruction in the limit of low-winding numbers.
}

\vspace{10pt}
\noindent\rule{\textwidth}{1pt}
\tableofcontents\thispagestyle{fancy}
\noindent\rule{\textwidth}{1pt}
\vspace{10pt}

\section{Introduction}
\label{sec:intro}

Quantities conserved by the dynamics, or more generally, invariants of
transformations, are essential to classify configurations and states
of matter.  They provide us with shortcuts to identify configurations
up to ``unessential'' deformations, by giving them a common label.
Examples of invariants range from basic symmetry operations
(\textit{e.g.} vector lengths for rotations), to complex problems
(knots, homotopy groups...), or appear unexpectedly in
some popular games (14-15 puzzle, Rubik's cube,
Solitaire...)\cite{Stewart}.
In the physical context, we
generally label the physical states according to the irreducible
representations of a symmetry group. However, the internal symmetries may be unknown or 
unexpected (\textit{e.g.} SO(4) for the hydrogen atom). 
The particle labels of a field theory at a given scale may thus be the result of non-obvious  invariants of the dynamics of more ``elementary'' degrees of freedom at a lower scale. 
Finding invariants in  the dynamics of simple degrees of freedom is therefore of particular interest, as it reveals some form of internal symmetry. 

Here we study the dynamics of a 
locally-constrained classical model, and provide the invariants that classify its sectors.  Locally-constrained models (also called ``vertex'' or ``ice''-type)
can be viewed as the strong-coupling
limit of physical models, in that the strong local
interactions are replaced by hard on-site constraints.
Among them, there are examples where
natural dynamical transformations lack
ergodicity\cite{Huse,Moore,Moharinv,Mohar,Normand,dimer,Freedman,Cepas}.
This is  the case of the
 3-coloring
model, which we consider here.   In this model, a hard local constraint prevents neighboring
edges to have the same color among three, on a regular hexagonal lattice (see Fig.~\ref{fig1}
for an example). It was originally introduced as an exactly-solvable model of ice-type where the entropy could be explicitly calculated\cite{Baxter}. The model was argued later to describe some magnetic materials, having the geometry of the kagome lattice \cite{kagome}, where the
three colors are three spin directions 
that are forced to be 
different by a strong local antiferromagnetic
interaction \cite{Huse,Chandra}. As a simple model, it may have various
other physical applications,  from a minimal
description of glasses \cite{Chakraborty,Castelnovo,Cepasglasses}, to superconducting arrays\cite{Castelnovo}. The
degrees of freedom that are compatible with the constraints are loops
of alternating colors (see Fig.~\ref{fig2}), similar to the moves
Kempe introduced to swap colors in planar maps. When periodic boundary
conditions are used, this dynamics is known to be
nonergodic, for reasons not fully understood \cite{Huse,Moore,Moharinv,Mohar,Cepas}. The color
configurations can be put in equivalence classes, called Kempe
sectors, if they are connected by the dynamics, the number of which,
$n_K>1$ \cite{Moharinv,Mohar}, has been enumerated numerically on small random
cubic graphs\cite{Belcastro}, or regular hexagonal
clusters\cite{Moharinv,oddeven}.
An invariant has been found, allowing to distinguish odd from even
colorings \cite{Moharinv,Mohar,oddeven}. It does not exhaust the number of
sectors \cite{oddeven}, so that the classification of Kempe sectors is so far incomplete.

With periodic boundary conditions, closed loops are characterized by
winding numbers which count the number of times the loops wind around
the boundaries \cite{note}. The dynamics is twofold. First, small local loops locally
deform the winding loops, but preserve their topological winding numbers and
thus define homotopy classes (flux conservation). Second, the dynamics of the winding loops 
 do not conserve the topological numbers (flux insertion). Since they are
integer numbers, this results in a dynamics described by
transformations from integers to integers.  The question arises as to
whether this integer dynamics can reach any configuration of winding numbers.
In fact, it cannot and there are invariants, stable with system size, that classify
the Kempe sectors in terms of polynomials of the winding
numbers (which are not individually conserved). While many sectors remain undistinguished by these
invariants, we will further argue that there is no other invariant, \textit{i.e.} this set is complete. In fact, paths between these additional sectors
involve intermediate configurations that exist at larger sizes: these
sectors at a given size are isolated because of steric obstruction.

In addition to the stable classification of the configurations, these results show the existence of many disconnected sectors of flux lines at fixed size. 
In the
context of the ``slow'' dynamics of glasses, nonergodicity plays a central role. It serves as a basic
property to generate new timescales, since the reconnection of 
these sectors may be assured by slower dynamical
processes.  This is what happens here as a consequence of the steric obstruction. On the technical side, these questions are relevant for
Monte Carlo simulations where nonlocal moves are used to sample the
space of configurations \cite{Newman}. Invariants introduce a bias that
has to be overcome, either by giving up the periodic boundary
conditions \cite{Moore}, or by enriching the moves with that of ``stranded''
loops \cite{oddeven}.

The paper is organized as follows. In section \ref{direct}, we recall
the direct numerical construction of Kempe sectors on small
clusters by constructing the 3-colorings and the dynamics between them exhaustively. In section \ref{integerdynamics}, we define the winding loop
configurations and the effective dynamics that transform the winding numbers. By
iterating this dynamics, we construct the Kempe sectors of winding numbers for much
larger systems. We then construct three invariants of the
dynamics in section \ref{invariants}, stable at all system sizes. We argue in section~\ref{steric}
that this set of invariants is complete: the remaining sectors result
from steric constraints and are absent when such constraints are relaxed
(section~\ref{steric}).

\section{Kempe sectors by direct construction}
\label{direct}

\begin{figure}[h]
\begin{center} \psfig{file=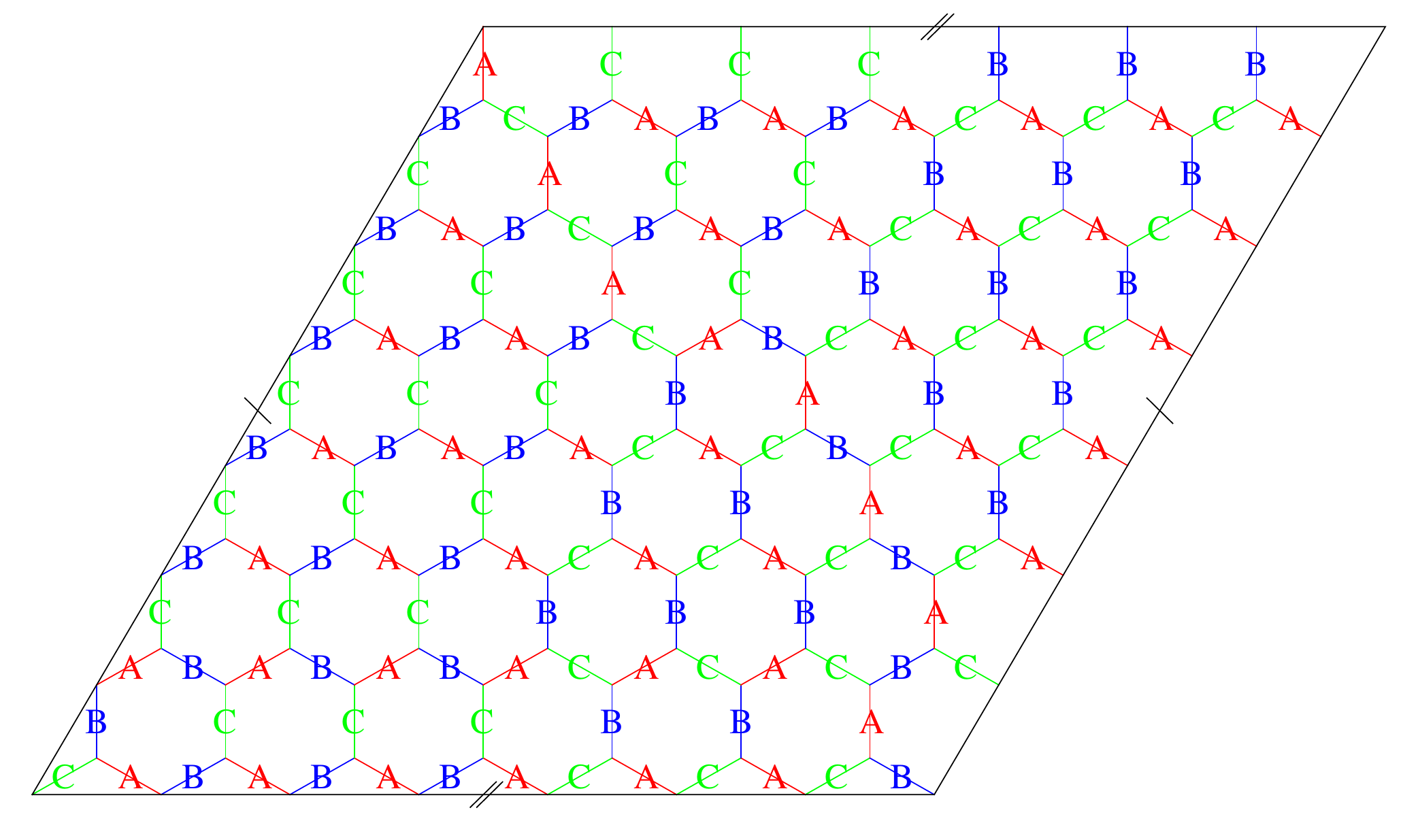,width=9.0cm,angle=-0} \hspace{1cm}
\end{center}
\caption{Example of a 3-coloring of the $N=3L^2$ edges of a hexagonal lattice of linear size $L=7$. Periodic boundary conditions are used to form a torus and define homotopy classes.}
\label{fig1}
\end{figure}

The model consists of coloring the edges of a regular hexagonal
lattice of linear size $L$ (and $N=3L^2$ edges) with three colors, \textit{e.g.} A, B, C, such that each vertex has three edges
colored with three different colors, see Fig.~\ref{fig1}. The number
of such 3-colorings has been calculated exactly by Baxter\cite{Baxter} and scales
as $\sim 1.1347^{N}$ in the thermodynamic limit. Periodic boundary conditions are
used, so the graph has the geometry of a torus and homotopy classes
exist.

\begin{figure}[h] \center
\psfig{file=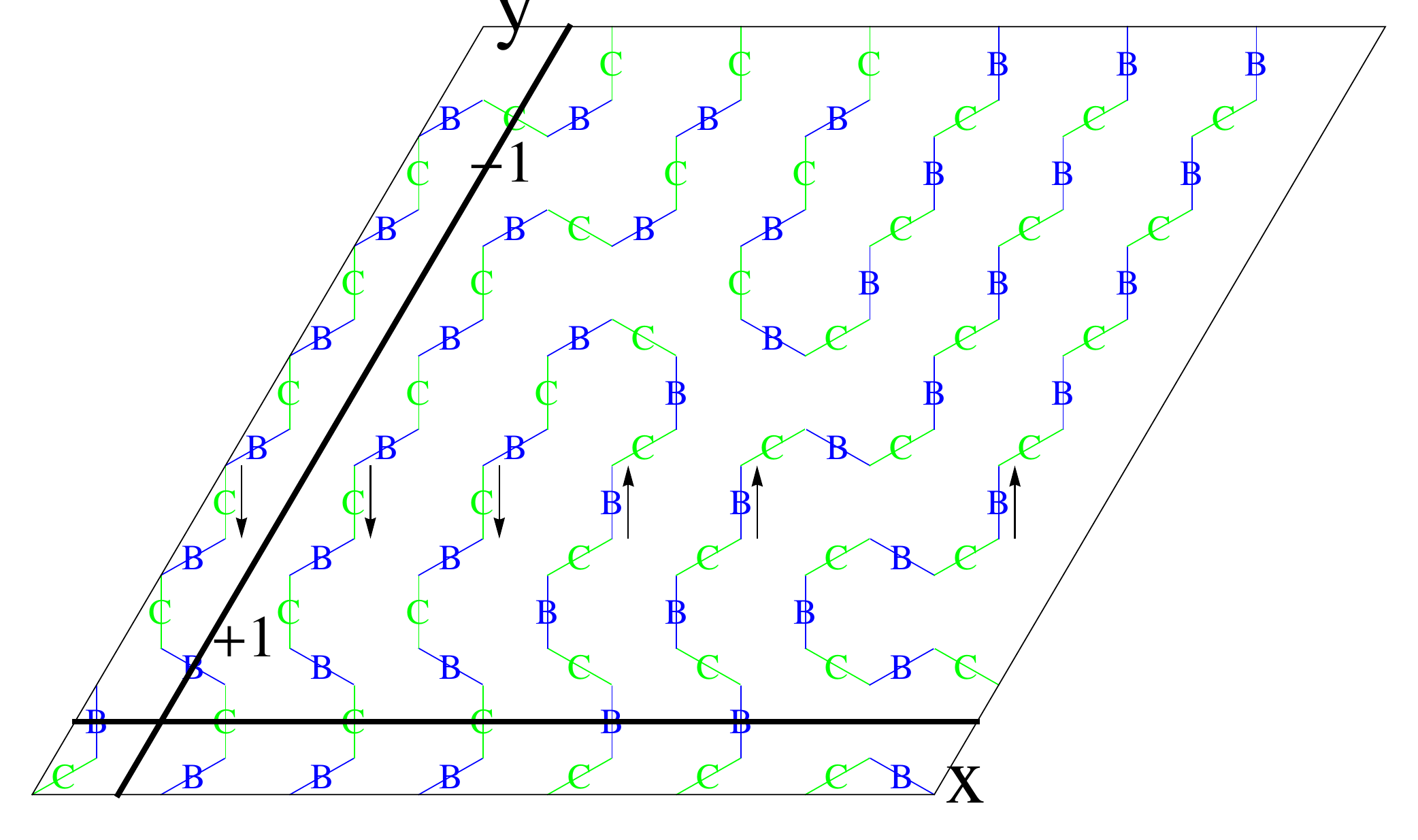,width=9.0cm,angle=-0} \hspace{1cm}
\caption{Same configuration as in Fig.~\ref{fig1} with red edges not shown, emphasizing closed self-avoiding B-C (blue-green) loops, which can be seen as flux lines through the $x$ and $y$ lines. Orientation of the loops is defined in Fig.~\ref{fig3}. There are three types of flux lines (not shown) which cross.}
\label{fig2}
\end{figure}
\begin{figure} \center \vspace{-.2cm}
\psfig{file=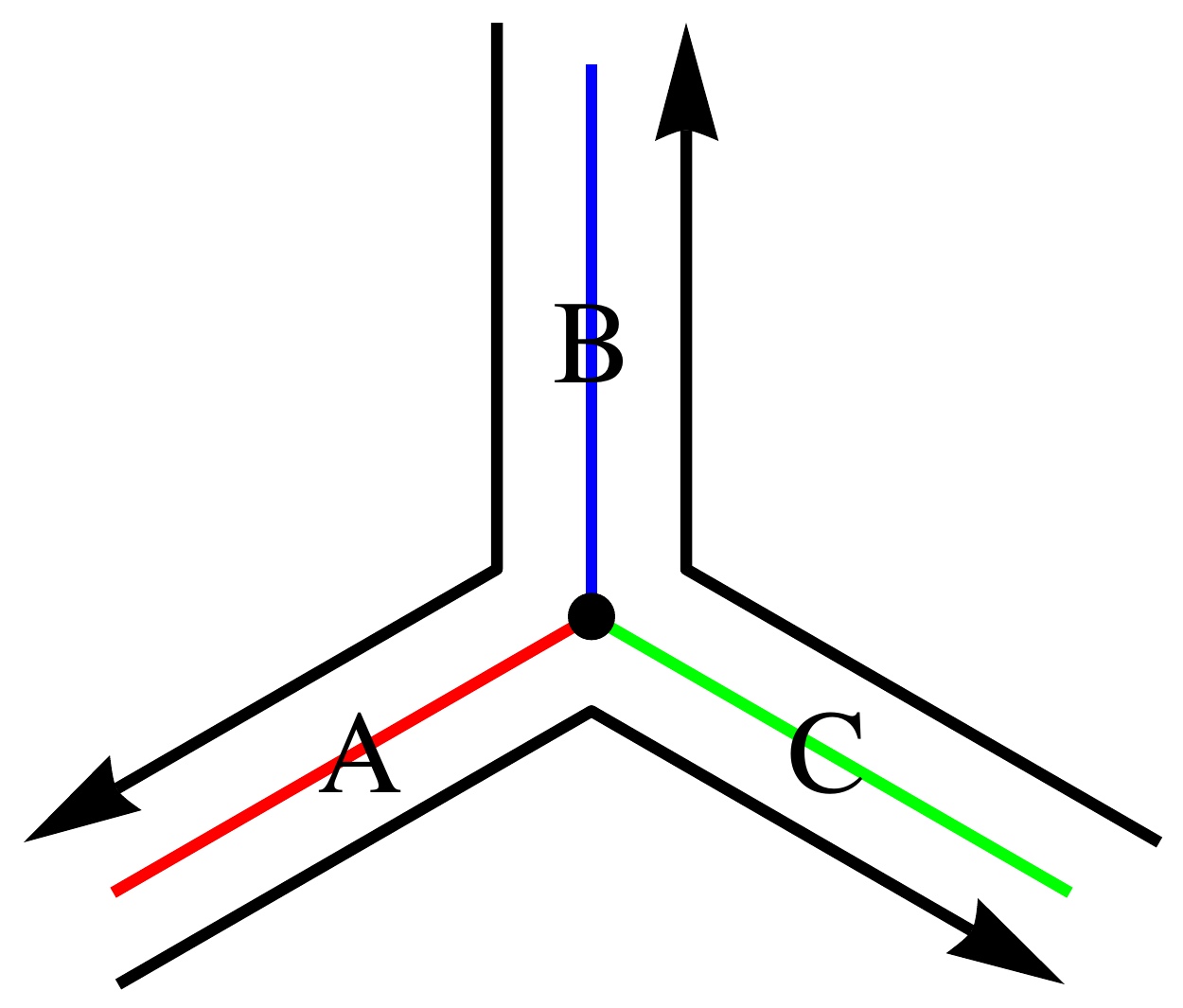,width=3.0cm,angle=-0} 
\caption{Definition of the orientation of the loops for a black vertex (the definition is opposite on white vertices).}
\label{fig3}
\end{figure}
The successive edges of two colors, say B and C (or A and B, or A
and C) form self-avoiding closed loops. All loops are fully-packed. In Fig.~\ref{fig2}, all B-C loops are
shown and edges colored A have been withdrawn. Exchanging the two colors of one of these 
loops gives a new
valid  3-coloring (all edges colored with three different colors). It is an example of Kempe move. There are different types of loops, $a$-type (B-C), $b$-type (A-C) and $c$-type (A-B). When a loop of one type flips, it reorganizes the other types of loops. The loops are also characterized by the number of times they wind across the two boundaries, the winding numbers $(p,q)$ (homotopy classes), the precise definition of which will be given in~\ref{windings}.

For completeness, we recall the construction of
Ref.~\cite{oddeven}. We first construct numerically all 3-colorings of a
cluster of size $L$. This procedure is obviously limited  by the
exponentially-growing number of states to small $L \lesssim 8$.  By iteration of the Kempe
moves, considering the motion of \textit{all} loops ($a$, $b$, $c$
type, winding and nonwinding), we find that not all states can be
connected but fall into closed disconnected sectors, called Kempe sectors.  The number of
Kempe sectors, $n_K$, depends on
$L$ and is given in table~\ref{tableKT}. It is
increasing with system size, but it is difficult to know whether it is
infinite in the thermodynamic limit. We also give the number of 3-colorings in each Kempe sector $Z_{i=1,\dots,n_K}$, their multiplicity, and the labels obtained in section~\ref{invariants}.

\begin{table}
\begin{center}
\begin{tabular}{r|r|p{7cm}}
  \hline \hline
  $L$  & $n_K$  & $Z_{i=1,\cdots,n_K}, I$  \\
  \hline
  2 & 1 & 24 (0) \\
  3 & 2 & 60 (0), 60 (1,-1,1)  \\
  4 & 2 & 240 (0), 1920 (1,-1,-1) \\
  5 & 3 & 35340 (0), 276 (1,-1,-1), 13800 (1,-1,1) \\
  6 & 4 & 2879856 (0), 307296 (1,-1,1), 19440($\times$2) (1,1,$\pm$1) \\
  7 & 2761 & 50683920 (0), 1140 (1,-1,1), 424598328 (1,-1,-1), 6 ($\times$2758) (1,1,$\pm$1) \\ 
  8 & 9 & 59539228896 (0), 54178583040 (1,-1,1), 14555136 (1,-1,-1), 28369152($\times$6) (1,1,$\pm$1)  \\ 
  \hline \hline
\end{tabular}
\end{center}
\caption{Number of Kempe sectors $n_K$ and number of 3-colorings in each sector $Z_{i=1,\dots,n_K}$ (sector multiplicity is indicated). The labels given in brackets are the purpose of the paper (defined in section~\ref{invariants}). For $L=7$ and $L=8$, they do not exhaust the number of sectors, see section~\ref{steric}.}
\label{tableKT}
\end{table}

The dynamics of nonlocal winding loops do not conserve the winding numbers, so that homotopy
classes are not preserved.  Nevertheless, they appear to be sets of
winding numbers that remain disconnected from other sets of winding
numbers.  Importantly, disconnected Kempe sectors having the same winding
numbers do not occur, in general. There is one exception in the range
of sizes available. For $L=7$, a large number of Kempe sectors is
found (table~\ref{tableKT}), all having the minimum number of 6
states. These are special 3-colorings, characterized by a single loop
of each type, of maximal size, $2N/3$ \cite{oddeven}. A translation of
these loops does not change the winding numbers but cannot be
connected, thus generating many special sectors.  Except for this
size, an homotopy class is found to be connected.

We study the main
obstruction, stable with system size, the one related to the winding numbers (keeping in mind that there may be finer obstructions for
special $L$), by setting up a dynamics for the winding
numbers themselves.

\section{Integer dynamics of winding numbers}
\label{integerdynamics}

The aim of this section is to set up the dynamics of the winding numbers, when nonlocal moves are allowed, \textit{i.e.} the set of transformations from integers to integers. We next study its ergodic properties, 
and what classes form.

\subsection{Definition of winding numbers}
\label{windings}
The total winding numbers of a configuration are defined by a triplet of vectors $(\bm{a},\bm{b},\bm{c})$,
\begin{eqnarray}
\bm{a} &=& \sum_l (a_{l}^x,a_{l}^y), \\
\bm{b} &=& \sum_l (b_{l}^x,b_{l}^y), \\
\bm{c} &=& \sum_l (c_{l}^x,c_{l}^y), 
\end{eqnarray}
where $\bm{a}$, $\bm{b}$, $\bm{c}$ are two-component vectors (we will define a third component below) corresponding to the three types of loops, $a \equiv$ B-C, $b \equiv$ C-A and $c \equiv$ A-B loops. The sum is over all loops $l$.
For instance, $a_l^{\alpha}$ is the winding number of the $l$ loop 
 across a cycle $\alpha=x$ or $y$ of the torus (Fig.~\ref{fig2}). They count the number of geometrical (signed)
 crossings. To compute them, we first orient the loops.  The  hexagonal
lattice is bipartite and we can define two sets of vertices, black and white, connected by the edges.
We then orient the loops
from B to A, A to C and C to B on black vertices (the opposite for white vertices), as shown in Fig.~\ref{fig3}. 
With this
convention of orientation, every B-colored edge that cuts the $y$-axis (and is part of a B-C loop), contributes to the winding number by $+1$ and every edge colored C contributes by $-1$. In the example of Fig.~\ref{fig2},
$a_l^y=1-1=0$ for the loop $l$ that crosses the $y$-axis in two places and $a_l^y=0$ for the other loops which do not cross the $y$-axis, so that $a_y=0$ (there is no flux through the $y$-axis).
Therefore, if we denote by $N_B^y$ and $N_C^y$ the number of edges along the $y$ line that are colored B and C, we get $a_y=N_B^y-N_C^y$.  Similarly, we define a set of numbers $N_i^{\alpha}$, $i=A,B,C$ and $\alpha=x,y$, so that
\begin{eqnarray}
a_x &=& N_B^x-N_C^x, \label{ax} \\
a_y &=& N_B^y-N_C^y, \\
b_x &=& N_C^x-N_A^x, \\
b_y &=& N_C^y-N_A^y, \\
c_x &=& N_A^x-N_B^x, \\
c_y &=& N_A^y-N_B^y. \label{cy}
\end{eqnarray}
Since we have the constraints on the number of colors, $0 \leq N_i^{\alpha} \leq L$, we get some constraints for the winding numbers,
\begin{equation}
  |a_{\alpha}| \leq L, \hspace{.6cm} |b_{\alpha}| \leq L, \hspace{.6cm} |c_{\alpha}| \leq L,
\end{equation}
where $\alpha=x,y$. It is also apparent that,
\begin{equation}
  \bm{a}+\bm{b}+\bm{c}=0.
  \label{sumabc}
\end{equation}
A  consequence is that it is not possible to have a single winding loop. Indeed, in this case, we would have (without loss of generality) $\bm{a}=(p,q) \neq 0$ and $\bm{b}=\bm{c}=0$ and the sum would not be zero. Note that a permutation of all colors B and C is an exchange of the winding numbers $\bm{b}$ and $\bm{c}$ with a global sign change, $\bm{a} \rightarrow -\bm{a}$ and $\bm{b} \rightarrow -\bm{c}$, $\bm{c} \rightarrow -\bm{b}$.

Since all of the $L$ edges cutting the lines $\alpha=x, y$ have some color, we have,
\begin{equation}
\sum_i N_i^{\alpha} = L
\label{sum1}
\end{equation}
for all $\alpha$.
It is convenient to define another set of numbers $N_i^z$, $i=A,B,C$, along a third line in the $z$ direction (the $y,z$ axis are at  $\pm 120^o$ with the $x$ axis). 
Indeed, the sum of the three numbers of a given color times $L$ is the total number of this color on the graph, \textit{e.g.} $L(N_A^x+N_A^y+N_A^z)=N/3=L^2$, and similarly for B and C:
\begin{equation}
\sum_{\alpha} N_i^{\alpha} =L
\label{sum2}
\end{equation}
for all $i=A,B,C$ ($N_A^z$ is not an independent number). Since $N_A^z$ must be positive and smaller than $L$ it gives three additional constraints: $N_i^x+N_i^y \leq L$. 

Given the two equations (\ref{sum1})  and (\ref{sum2}), we can invert Eqs.~(\ref{ax})-(\ref{cy}).
It is convenient to define a third component $a_z$ such that $a_x+a_y+a_z=0$ (and similar definitions for $b$ and $c$). We can write the relations between $(\bm{N}_A,\bm{N}_B,\bm{N}_C)$ and $(\bm{a},\bm{b},\bm{c})$ in the form,
\begin{eqnarray}
\bm{N}_A &=& \frac{L}{3} \mathbf{1} + \frac{1}{3} (\bm{c}-\bm{b}), \label{eq1} \\
\bm{N}_B &=& \frac{L}{3} \mathbf{1} + \frac{1}{3} (\bm{a}-\bm{c}), \label{eq2} \\
\bm{N}_C &=& \frac{L}{3} \mathbf{1} + \frac{1}{3} (\bm{b}-\bm{a}), \label{eq3}
\end{eqnarray}
where $\mathbf{1}=(1,1,1)$ and all vectors have now three (non-independent) components. There is a central sector, called the ``0-sector'' characterized by the absence of winding loops, $(\bm{a},\bm{b},\bm{c})=0$, which exists when $L$ is a multiple of three. A finite $(\bm{a},\bm{b},\bm{c})$ can be seen as a deviation from this sector, but the integers must be chosen such that the $\bm{N}_{\alpha}$ themselves are integers, \textit{i.e.} $\bm{b}-\bm{c}$, $\bm{c}-\bm{a}$, and $\bm{a}-\bm{b} \equiv L$(mod 3). To have the same set of integers  $(\bm{a},\bm{b},\bm{c})$ at $L$ and $L_1$, we must have $L_1 \equiv L (\mbox{mod} 3)$. In particular, changing the sign of $(\bm{a},\bm{b},\bm{c}) \rightarrow - (\bm{a},\bm{b},\bm{c})$ is possible only when $L$ is a multiple of three, since $\bm{b}-\bm{c}$ or $\bm{c}-\bm{b}$ are both multiple of three, in this case.

We can define a norm for the winding numbers,
\begin{equation}
n^2= \frac{1}{6} (|\bm{a}|^2+|\bm{b}|^2+|\bm{c}|^2), \label{normdef}
\end{equation}
where $|\bm{a}|^2=a_x^2+a_y^2+a_z^2$. It has a minimum at $n=0$ in the ``0-sector'' and a maximum at $n=L$.
The conditions $|\bm{a}| \ll L$, $|\bm{b}| \ll L$, and $|\bm{c}| \ll L$ (or $n \ll L$) define the limit of low winding numbers. It can be seen as a \textit{dilute} limit, since low-winding number loops may occupy a number of sites of order $L$, \textit{i.e.} a \textit{vanishing} fraction of the lattice sites. On the contrary, when some winding numbers are a fraction of $L$, the loops occupy a \textit{finite} fraction of the lattice sites (\textit{finite-density} configurations).

A color configuration is thus reduced to a set of nine winding numbers $(\bm{a},\bm{b},\bm{c})$ or, equivalently, $(\bm{N}_A,\bm{N}_B,\bm{N}_C)$. Only four of them are independent (given Eq.~(\ref{sumabc}) and the definition of the $z$-components). A color configuration should 
satisfy the ``box'' constraints,
\begin{equation}
  0 \leq N_i^{\alpha}\leq L, \label{c0}
\end{equation}
where $i=A,B,C$ and $\alpha=x,y,z$.
  Given these constraints, the actual number of winding number configurations is a fraction of $L^4$, and, of course, much smaller than the number of color configurations.

\subsection{Setting up the transformations of winding numbers}

The dynamics we consider consists of exchanging the two colors A-B, B-C or C-A along any closed loop $l$. This is the simplest dynamics that preserves the constraints.

We now construct a ``coarse-grained'' dynamics in the much smaller
space of winding number configurations. A nonwinding loop modifies locally the shapes of the winding loops but
does not change the set of numbers $(\bm{a},\bm{b},\bm{c})$.

Let us consider a winding loop of type $a$ (B-C) 
with $\bm{a}_l=(1,0)$. The loop winds around the $x$-axis in the $y$ direction.
It intersects the $x$-axis in a site with
color B, given the convention of orientation. The flip of this loop changes the color on the $x$-axis from
B to C. $a_x=N_B^x-N_C^x$ becomes $a_x-2$. Simultaneously,
$b_x=N_C^x-N_A^x=b_x+1$ and $c_x=N_A^x-N_B^x=c_x+1$. More generally,
if the winding number is $a_x=w>0$, there are $w$ sites along the
$x$-axis which are in color B and which change to C after the flip, so
that $a_x =N_B^x-N_C^x \rightarrow a_x+2w=a_x+2a_l^x$, and $b_x$ or
$c_x$ change according to $b_x=N_C^x-N_A^x=b_x-w=b_x-a_l^x$.  The same
reasoning applies for the $y$ component, so that for a general flip of
a loop with winding numbers $\bm{a}_l \rightarrow -\bm{a}_l$, we
have the transformation law,
\begin{eqnarray}
& \bm{a}' & = \bm{a} - 2 \bm{a}_l, \\
& \bm{b}' & = \bm{b}+\bm{a}_l, \\
&\bm{c}'& = \bm{c}+\bm{a}_l.
\end{eqnarray}
$\bm{a}_l$ is parallel to $\bm{a}$, because all winding loops
of a given type are self-avoiding (if $\bm{a}_l$ were
not parallel to $\bm{a}$, we would have $\bm{a} \times \bm{a}_l \neq 0$ and they would cross). Let us define a \textit{primitive}
integer vector $\hat{\bm{a}}=(\hat{a}_x,\hat{a}_y)$ parallel to $\bm{a}$ but with the
smallest integer coefficients, 
\begin{equation}
\hat{\bm{a}}=\frac{\bm{a}}{\mbox{gcd}(a_x,a_y)},
\end{equation} 
where $\mbox{gcd}(x,y)>0$ is the greatest common factor of $x$ and $y$. The primitive vector $\hat{\bm{a}}$ characterizes the elementary winding loops, and gcd$(a_x,a_y)$ is the number of such loops. We can choose $\bm{a}_l=k \hat{\bm{a}}$ with $k$ a positive or negative integer:
\begin{equation}
  (\bm{a}', \bm{b}', \bm{c}')= (\bm{a} + 2k \hat{\bm{a}},\bm{b} - k\hat{\bm{a}} ,\bm{c} - k\hat{\bm{a}}).
 \end{equation}
 In terms of the number of colors, we have
$(\bm{N}_A',\bm{N}_B',\bm{N}_C') = (\bm{N}_A, \bm{N}_B + k \hat{\bm{a}} , \bm{N}_C - k \hat{\bm{a}})$, which makes explicit that a flip of a $a$-type loop exchanges the B and C colors along the different cycles.
There are two other similar transformations on A-B or A-C loops.

In summary, the dynamics consists of three transformations applied on the set of winding numbers,
\begin{eqnarray}
T_a(k):  (\bm{a}',\bm{b}',\bm{c}')   &=&  (\bm{a} + 2 k\hat{\bm{a}}, \bm{b} -  k\hat{\bm{a}}, \bm{c}- k\hat{\bm{a}}),  \label{dl1} \\
T_b(k):  (\bm{a}',\bm{b}',\bm{c}')   &=&  (\bm{a} -  k\hat{\bm{b}}, \bm{b} + 2 k \hat{\bm{b}}, \bm{c}- k \hat{\bm{b}}), \label{dl2}  \\
T_c(k):  (\bm{a}',\bm{b}',\bm{c}')   &=&  (\bm{a} -  k\hat{\bm{c}}, \bm{b} -  k\hat{\bm{c}} , \bm{c}+2k \hat{\bm{c}}),
\label{dl3}
\end{eqnarray}
where $k$ is an arbitrary positive or negative integer. The transformations are such that $\bm{a}'+\bm{b}'+\bm{c}'=0$.
Since, among the nine integers $(\bm{a},\bm{b},\bm{c})$, only four are independent, the transformations are
 maps from Z$^4$ onto Z$^4$. 

\subsubsection{First example: permutation of colors}

Consider a transformation $T_a(k_a)$ with $k_a=-$gcd$(a_x,a_y)$, \textit{i.e.} $k_a \bm{\hat{a}}=-\bm{a}$:
\begin{equation}
T_a(k_a): (\bm{a}',\bm{b}',\bm{c}')    =  (-\bm{a},-\bm{c},  -\bm{b})
\end{equation}
since $\bm{a}+\bm{b}+\bm{c}=0$.
From Eqs.~(\ref{eq1})-(\ref{eq3}), we see that
$ (\bm{N}_A',\bm{N}_B',\bm{N}_C')= (\bm{N}_A,\bm{N}_C,\bm{N}_B)$, which is a permutation of B and C.
We have six such permutations, generated by $P_i\equiv T_{i}(k_{i})$, $i=a,b,c$,
\begin{eqnarray}
 (\bm{a},\bm{b},\bm{c}),
 -(\bm{a},\bm{c},\bm{b}), -(\bm{c},\bm{b},\bm{a}), -(\bm{b},\bm{a},\bm{c}), (\bm{b},\bm{c},\bm{a}), (\bm{c},\bm{a},\bm{b}). \nonumber 
  \end{eqnarray}
Note that the triplet $(\bm{a},\bm{b},\bm{c})$ acquires a sign according to whether the permutation is even or odd (it transforms like the signature representation). 

\subsubsection{Second example: insertion of flux}

\begin{figure} \center \vspace{-.2cm}
\psfig{file=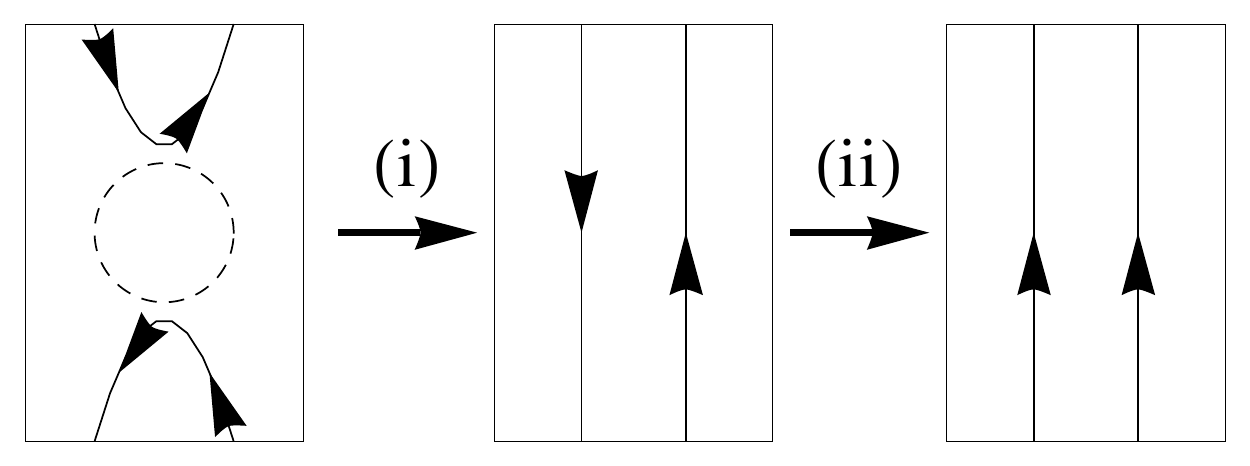,width=7.0cm,angle=-0} 
\caption{Creation of two opposite winding loops after the flip of the dashed loop (i) and insertion of flux after the flip of the winding loop (ii).}
\label{fig3l}
\end{figure}

Suppose that gcd$(a_x,a_y)=1$ and consider,
\begin{eqnarray}
T_a(1): (\bm{a}', \bm{b}', \bm{c}')=(  3\bm{a}, \bm{b}-\bm{a}, \bm{c}-\bm{a})
\end{eqnarray}
which describes a new configuration with three parallel winding loops of type $a$, since gcd$(a_x',a_y')=3$. 
It becomes in terms of number of colors,
$(\bm{N}_A', \bm{N}_B', \bm{N}_C') = (\bm{N}_A, 2 \bm{N}_B -\bm{N}_C, 2 \bm{N}_C -\bm{N}_B)$.
Physically, it corresponds to the insertion of flux after creation of two opposite winding loops (Fig.~\ref{fig3l}). They must be parallel to the existing ones (\textit{i.e.} noncrossing), which is taken care of, since the increase is along the direction of $\bm{a}$. There must be room for them. The second condition leads to some constraints (see below).

\subsubsection{Composition of transformations}

The transformations do not commute in general. We have some useful relations,
\begin{eqnarray}
  T_{i}(-1).T_i(1)&=& \mbox{Id}, \\
  T_{i}(1).T_i(1)&=&T_{i}(2), \\
  T_{i}(1).T_i(-1)&=&T_{i}(-2), 
\end{eqnarray}
where Id is the identity and $i=a,b,c$. This allows to restrict to $k=\pm 1$  and iterate.

\subsubsection{Constraints}

The transformations Eqs.~(\ref{dl1})-(\ref{dl3}) are not always possible because they may violate the space constraints:
\begin{eqnarray}
  0 \leq (N_i^{\alpha})' \leq L \label{c1}
\end{eqnarray}
where the prime denotes the numbers after the transformation. If one of these conditions is violated, the transformation is forbidden.

The transformations Eqs.~(\ref{dl1})-(\ref{dl3}) together with the constraints Eq.~(\ref{c1}) constitute the ``laws of dynamics'' for the winding numbers. We have ``coarse-grained'' the dynamics by eliminating the effects of the dynamics of the local loops.

\subsection{Numerical construction of Kempe sectors of winding numbers}
\label{numconstruction}

We start by constructing all initial sets of integers $(\bm{a},\bm{b},\bm{c})$ for a given $L$ satisfying the basic constraints [Eqs.~(\ref{eq1})-(\ref{eq3}),(\ref{c0})]. There are less than $L^4$ configurations, but more that the number of allowed winding-number classes in the coloring problem, so some of them are unphysical. While in the continuum limit we would expect all possible winding numbers, here the discreteness of the lattice gives additional constraints. Similarly, all winding numbers should be realized on the lattice but in the dilute limit\cite{notew}.
Away from this limit, the set of configurations is enlarged, but we can still get some qualitative insights.

\begin{figure}[h] \center \hspace{1cm}
\psfig{file=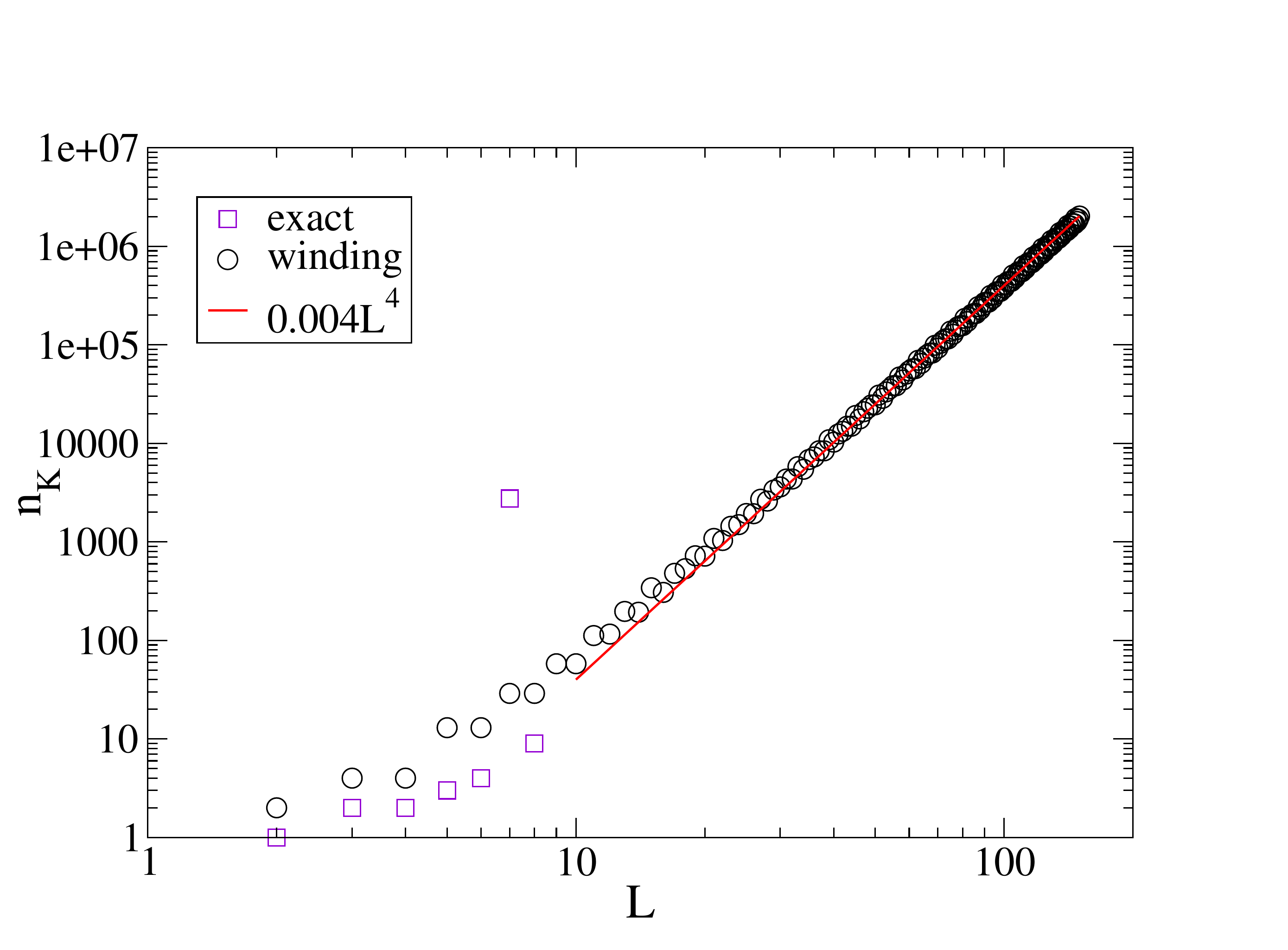,width=11.0cm,angle=-0} 
\caption{Number of Kempe sectors as a function of linear system size for the coloring problem (squares) and derived by assuming all configurations of winding numbers (circles). They form a finite fraction of the total number of configurations.}
\label{fig4}
\end{figure}
\begin{figure}[h] \center \hspace{1cm}
\psfig{file=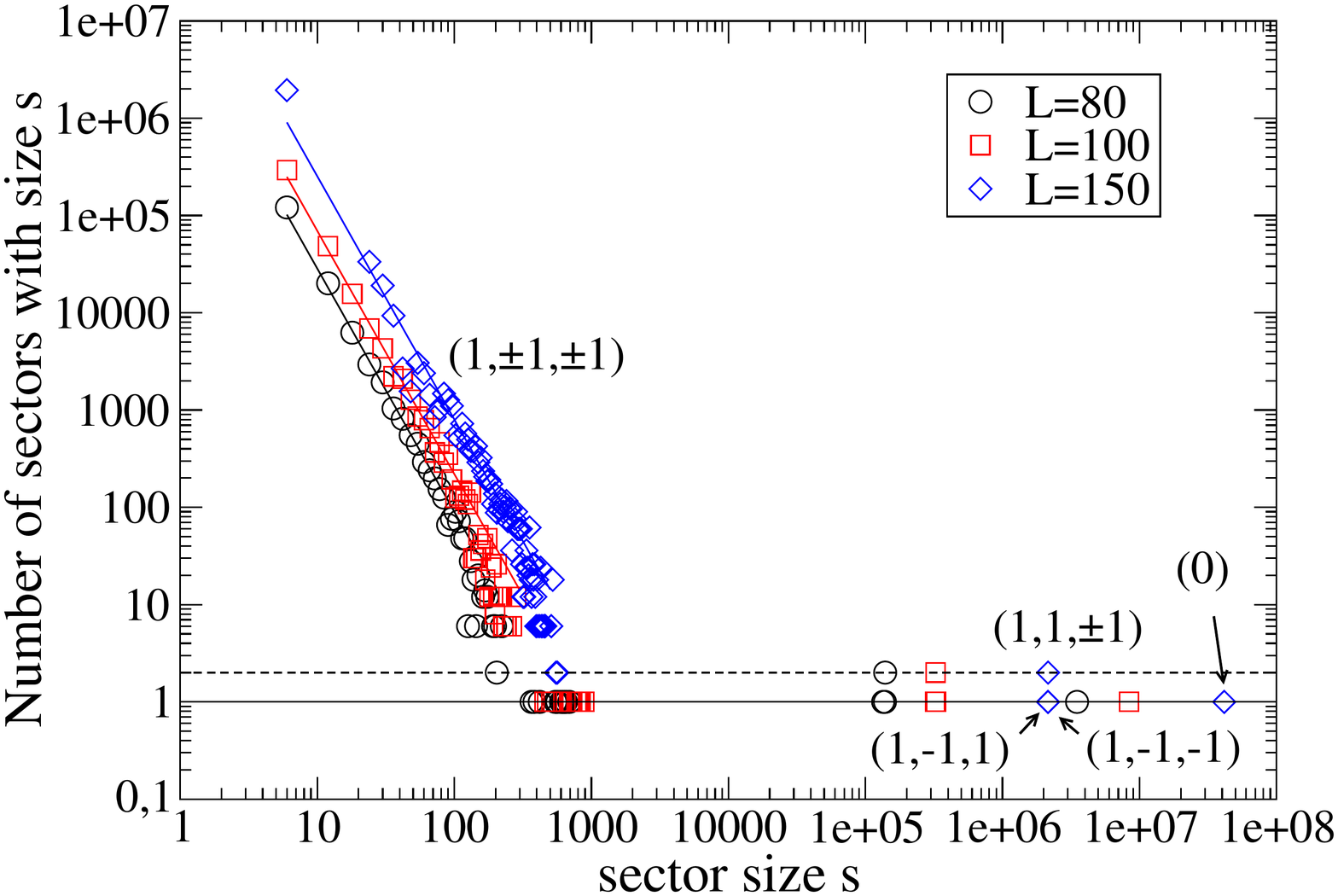,width=11.0cm,angle=-0} 
\caption{Distribution of sizes of Kempe sectors (log-log scale). It has a large number of small sectors (smaller than $\sim L$) and five sectors around macroscopic values (the dashed line indicates two sectors). The five larger sectors are labeled by the invariants found in section~\ref{invariants} (shown for $L=150$, the two with $(1,-1,\pm 1)$ are almost degenerate).}
\label{fig5}
\end{figure}

We have numerically iterated the dynamical transformations, Eqs.~(\ref{dl1})-(\ref{dl3}), from the initial configurations until the sectors close.
In Fig.~\ref{fig4}, we show the number of Kempe sectors of winding numbers as a function
of system size $L$. At small $L$, we can compare with that of table~\ref{tableKT}:
\begin{center}
\begin{tabular}{r|rrrrrrr}
  $L$             & 2 & 3 & 4 & 5 & 6 & 7 & 8 \\ \hline 
   $n_K$ (exact)   & 1 & 2 & 2 & 3 & 4 & 2761 & 9 \\ \hline
 $n_K$ (winding) & 2 & 4 & 4 & 13 & 13 & 29 & 29 
\end{tabular}
\end{center}
The differences between the two lines have two origins. First, some of the sets of integers $(\bm{a},\bm{b},\bm{c})$ that do not exist in the coloring
problem form additional Kempe sectors by themselves.
Second, for special $L$ ($L=7$), a few $(\bm{a},\bm{b},\bm{c})$
 classes are themselves split into subsectors (as discussed in section~\ref{direct}), giving additional Kempe sectors.

 We now have access to much larger system sizes and we can study the number of sectors of the winding number configurations.
 At large $L$,
$n_K$ converges to 0.004$L^4$ (solid line), which is a finite fraction
of all configurations. We emphasize that it is not the number of Kempe sectors of the original
problem and we do not know how many configurations among them are unphysical. A first point, however,  is that the winding number configurations themselves form a large number of 
disconnected sectors.

In Fig.~\ref{fig5}, we show the distribution of the number $s$ of configurations within the Kempe sectors. It has three peaks: one peak around zero with a small number of states in each sector, and two peaks with a macroscopic number of states (a fraction of $L^4$) split into five sectors (among them two have exactly the same number of states, and two others differ by a few percents).  The distribution around zero is close to a power law (shown with lines in Fig.~\ref{fig5}), $\sim s^{-2.5}$. Given the divergence of its integral at zero, the total number of sectors $n_K$ in Fig.~\ref{fig4}, is dominated by the smallest sectors, those with only six states.

In summary, we observe two types of Kempe sectors, those with $s \ll L^4$ and five large sectors. 
We will now explain that the sectors can be labeled by three invariants that are polynomials of the winding numbers (section~\ref{invariants}), and that the small sectors result from the steric constraints imposed by Eqs.~(\ref{c0}) (section~\ref{steric}).

\section{Stable invariants}
\label{invariants}

In this section, we show some conservation laws, \textit{i.e.} explicit functions of $( \bm{a}, \bm{b}, \bm{c})$ that are conserved by the integer dynamics and take different values in the different sectors. They are stable invariants, valid for all system sizes.

\subsection{Conservation of the parity of the number of line crossings}

We define 
\begin{equation}
  \bm{a} \times  \bm{b}=a_x b_y-a_y b_x, 
  \end{equation}
which is a one-component integer (the other components, such as $a_zb_x-a_xb_z=-(a_x+a_y)b_x+a_x(b_x+b_y)=a_xb_y-a_yb_x$, are equal to it). It counts the number of geometrical (signed) crossings of $a$-type loops with $b$-type loops.

Its symmetric version is
\begin{equation}
\chi=\frac{1}{3}(\bm{a} \times  \bm{b} +  \bm{b}\times  \bm{c} +  \bm{c}\times  \bm{a}) \label{chidef}
  \end{equation}
since, $   \bm{a} \times  \bm{b}=  \bm{b} \times  \bm{c} =  \bm{c} \times  \bm{a}$ results from $ \bm{a}+ \bm{b}+ \bm{c}=0$. So, for instance, if there are no A-B winding loops, $ \bm{c}=0$ and $\chi=0$: B-C and A-C winding loops (if they exist) are parallel in the geometrical sense.

In the transformation Eq.~(\ref{dl1}),
\begin{equation}  \bm{a}' \times  \bm{b}'= \bm{a} \times  \bm{b} + 2k \hat{ \bm{a}} \times  \bm{b} \end{equation}
has its parity conserved. The same argument applies for the two other transformations Eq.~(\ref{dl2}) and (\ref{dl3}). Therefore,  the quantity,
\begin{equation}
I_2^a=\chi \mbox{mod} 2 
  \end{equation}
is conserved by the dynamics.
$I_2^a$ defines two sectors, characterized by an even or odd number of geometrical crossings of the winding loops.
The sector of even parity is \textit{connected} and the proof (given in Appendix~\ref{connected}) consists of showing that all configurations are connected to a target configuration. The sector of odd parity is \textit{not} connected and we provide below further invariants in the odd sector.

\subsubsection{Relationship with odd/even chirality invariant}

\begin{figure}[h] \center 
\psfig{file=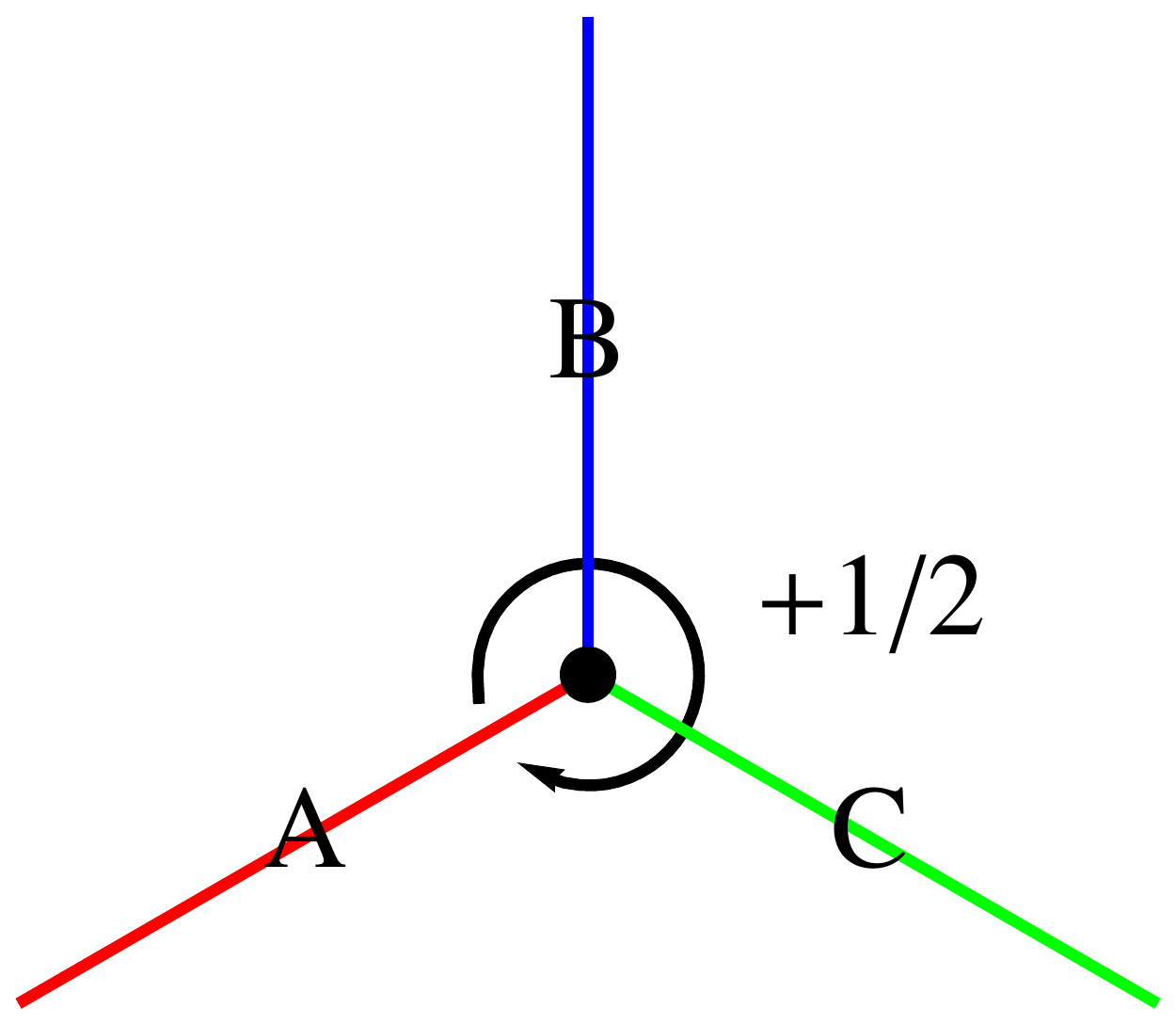,width=3.0cm,angle=-0} \hspace{1cm} 
\psfig{file=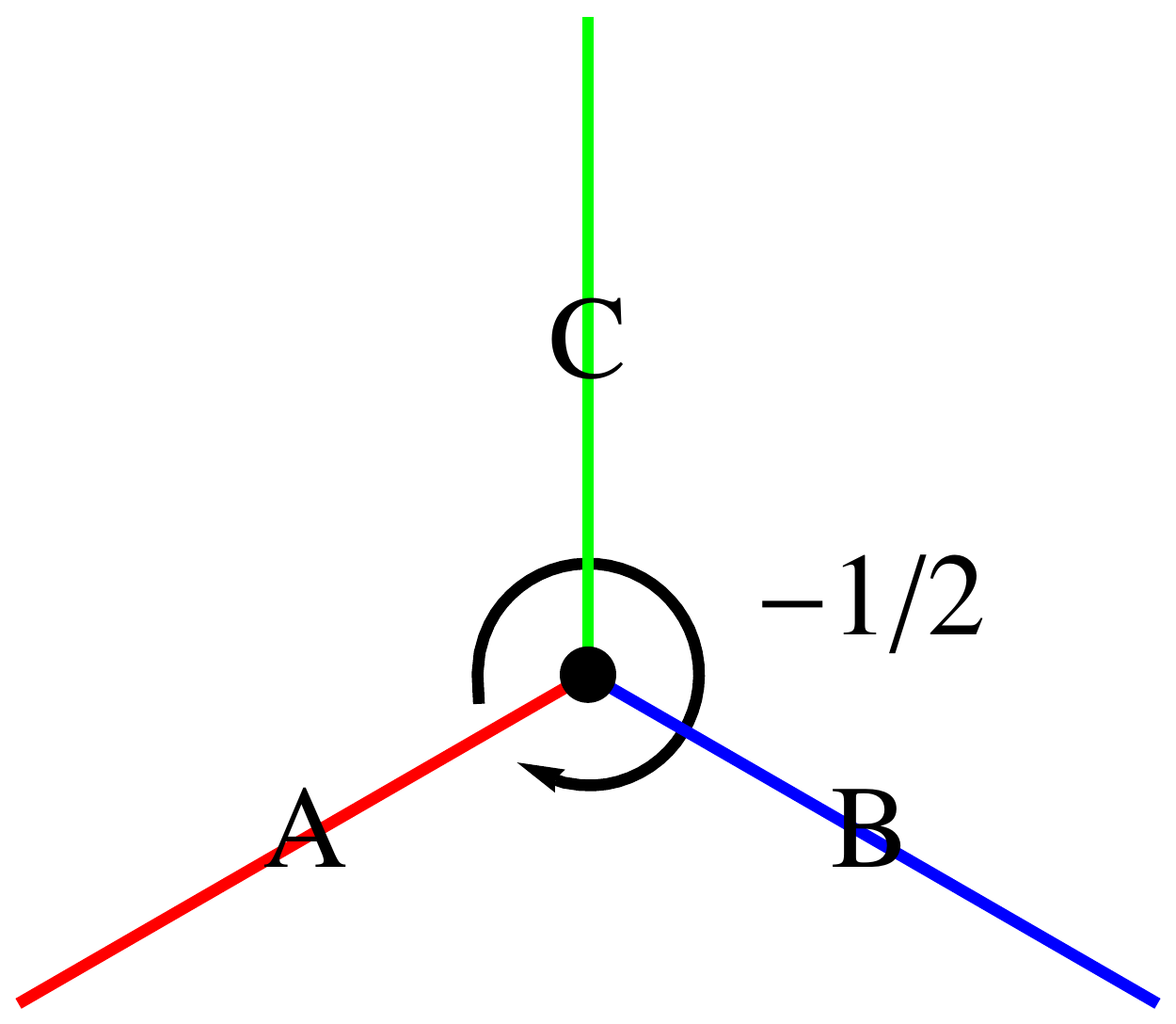,width=3.0cm,angle=-0} 
\caption{Definition of the chirality of a vertex.}
\label{fig6}
\end{figure}

We now explain why it is the same invariant as the parity of the chirality, previously defined \cite{oddeven}. 
We define the chirality at a vertex by choosing an orientation  and count $+1/2$ for an ABC ordering and $-1/2$ for ACB (see Fig.~\ref{fig6}, the same definition applies for black and white vertices). The chirality of a color configuration is the sum over the $2L^2$ vertices and is an odd or even integer number between $-L^2$ and $+L^2$. Its parity is conserved by the dynamics \cite{oddeven}. 
\begin{figure}[h] \center
\psfig{file=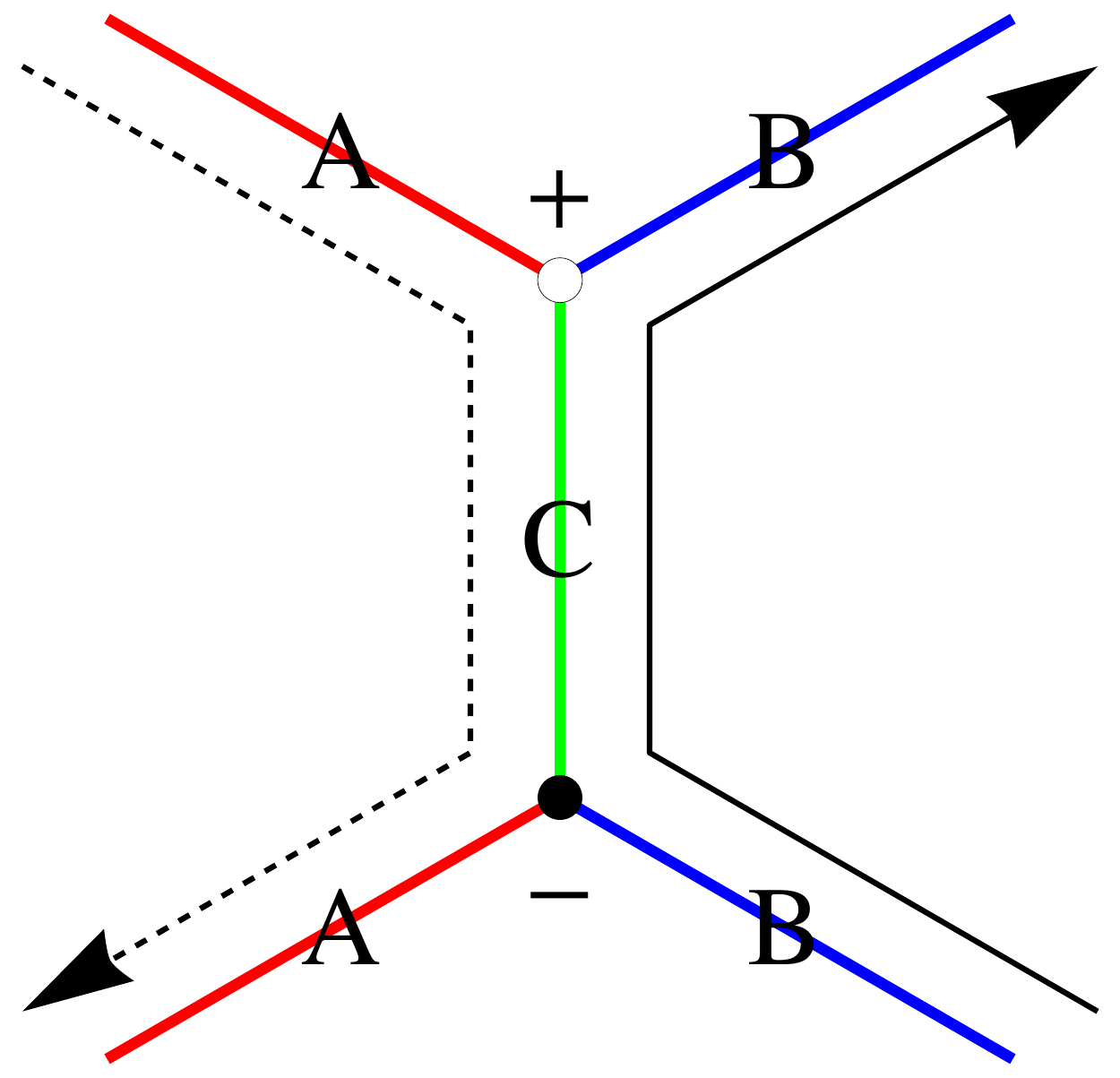,width=3cm,angle=-0} \hspace{.3cm} 
\psfig{file=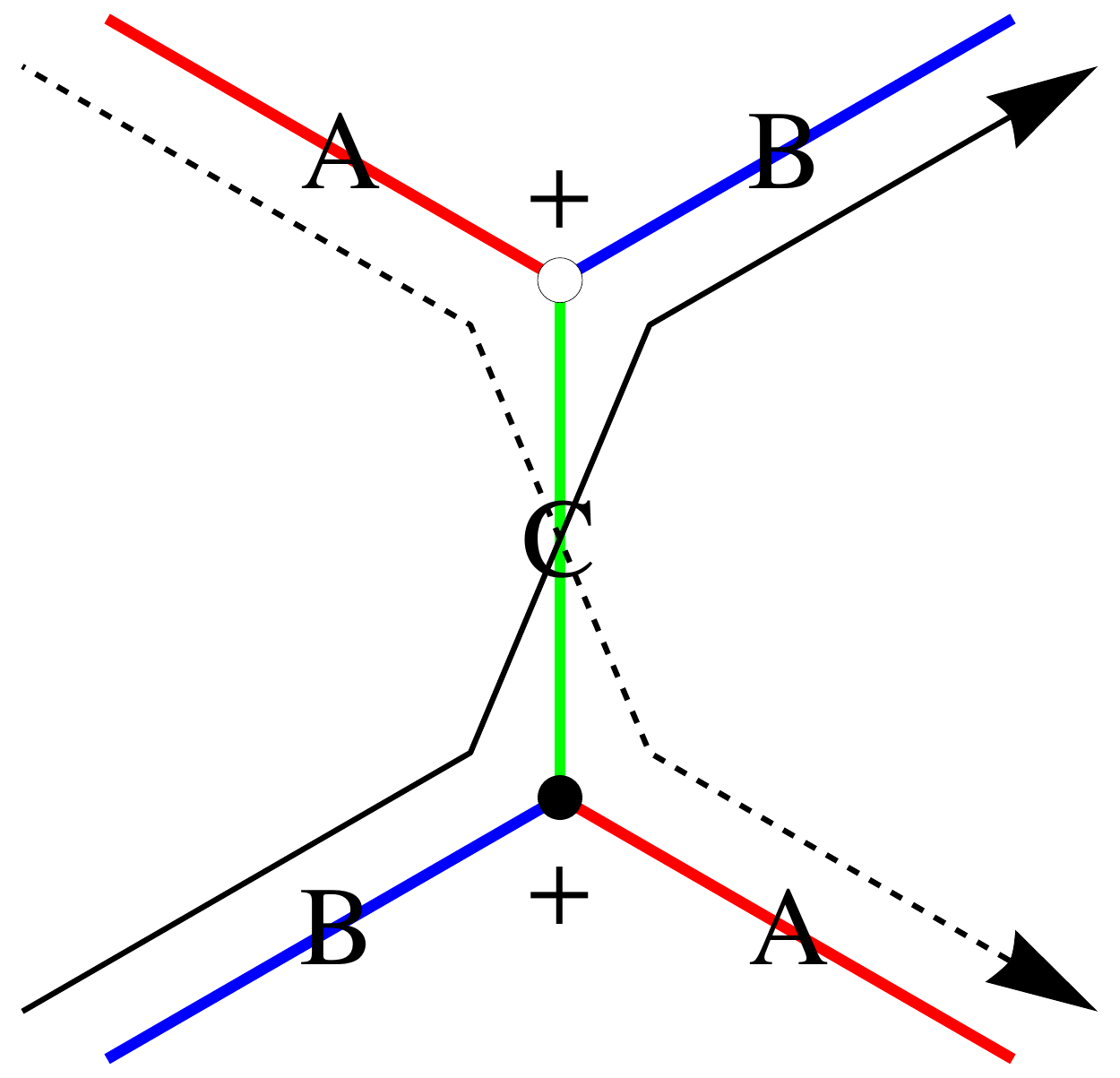,width=3cm,angle=-0} \hspace{.3cm} 
\psfig{file=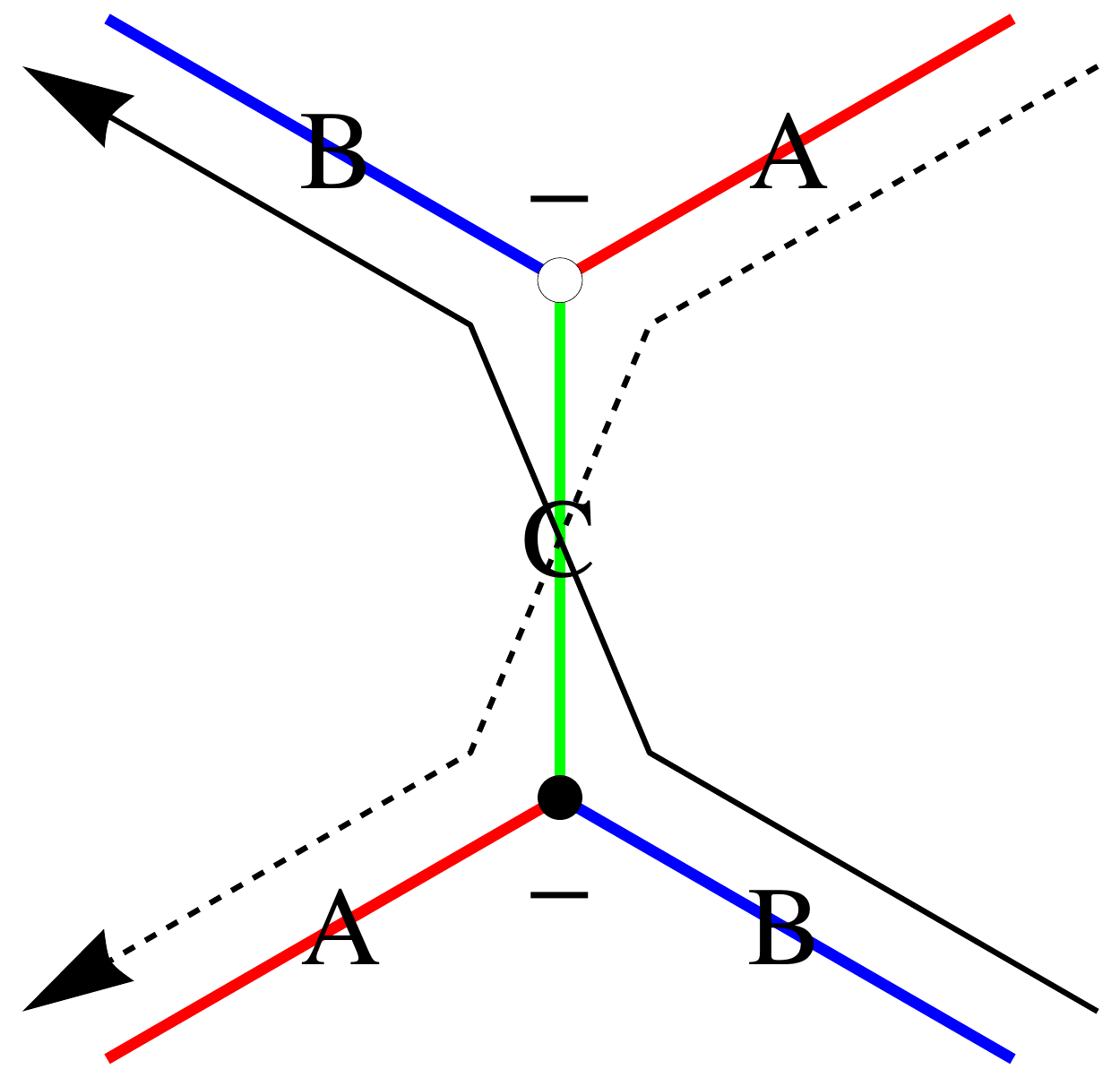,width=3cm,angle=-0} \\
\vspace{.3cm}
\psfig{file=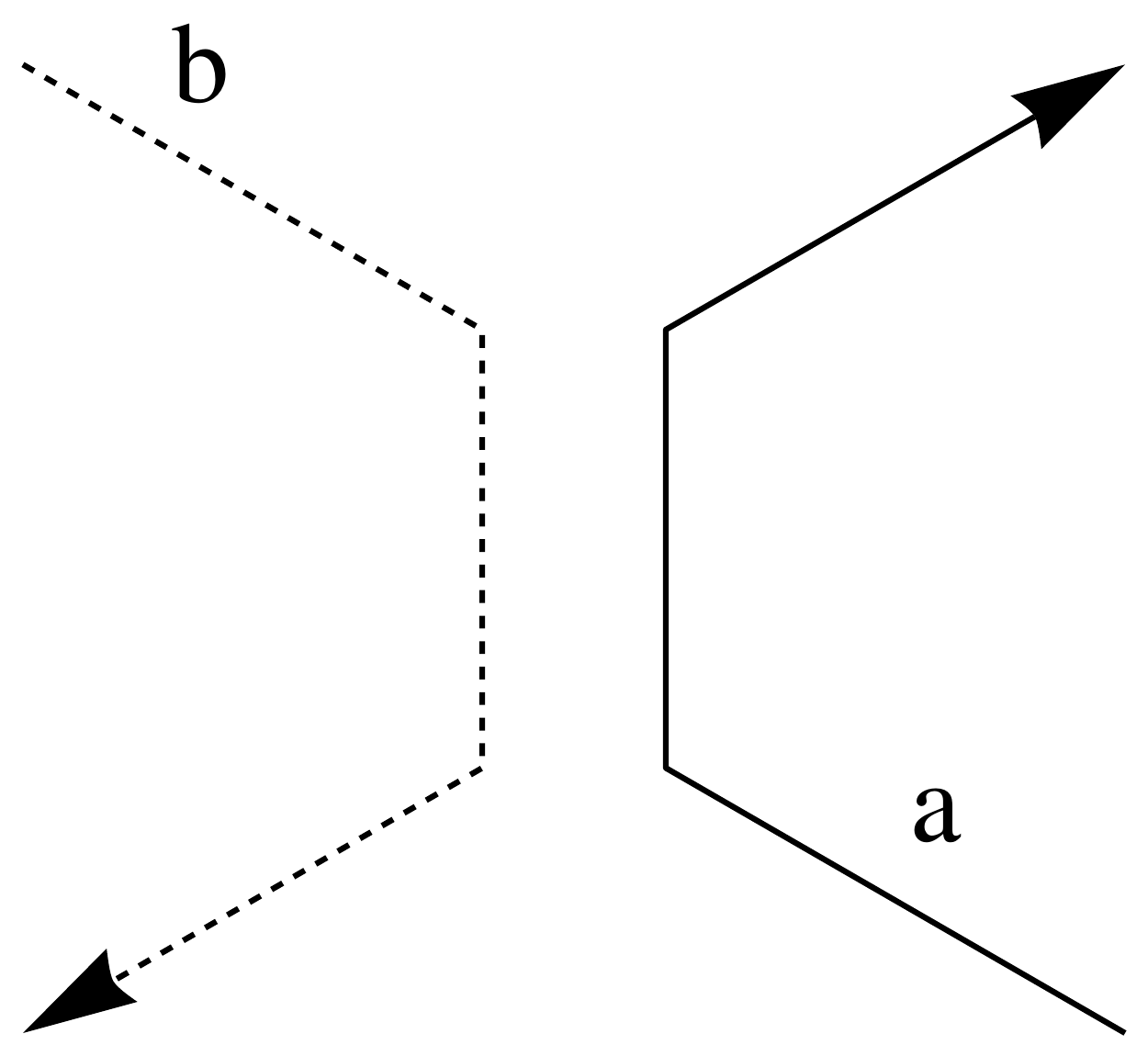,width=3cm,angle=-0} \hspace{.3cm} 
\psfig{file=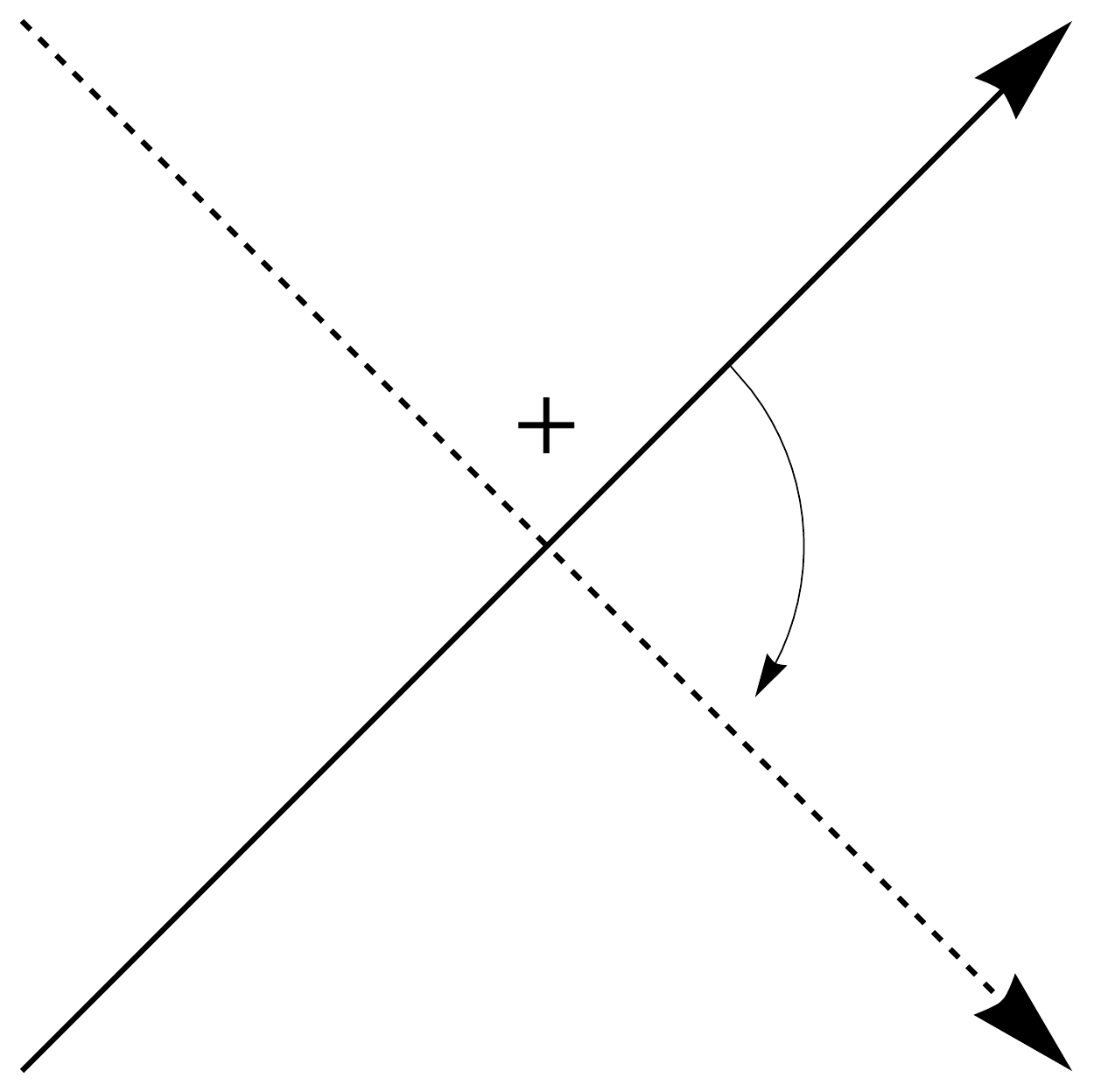,width=3cm,angle=-0} \hspace{.3cm} 
\psfig{file=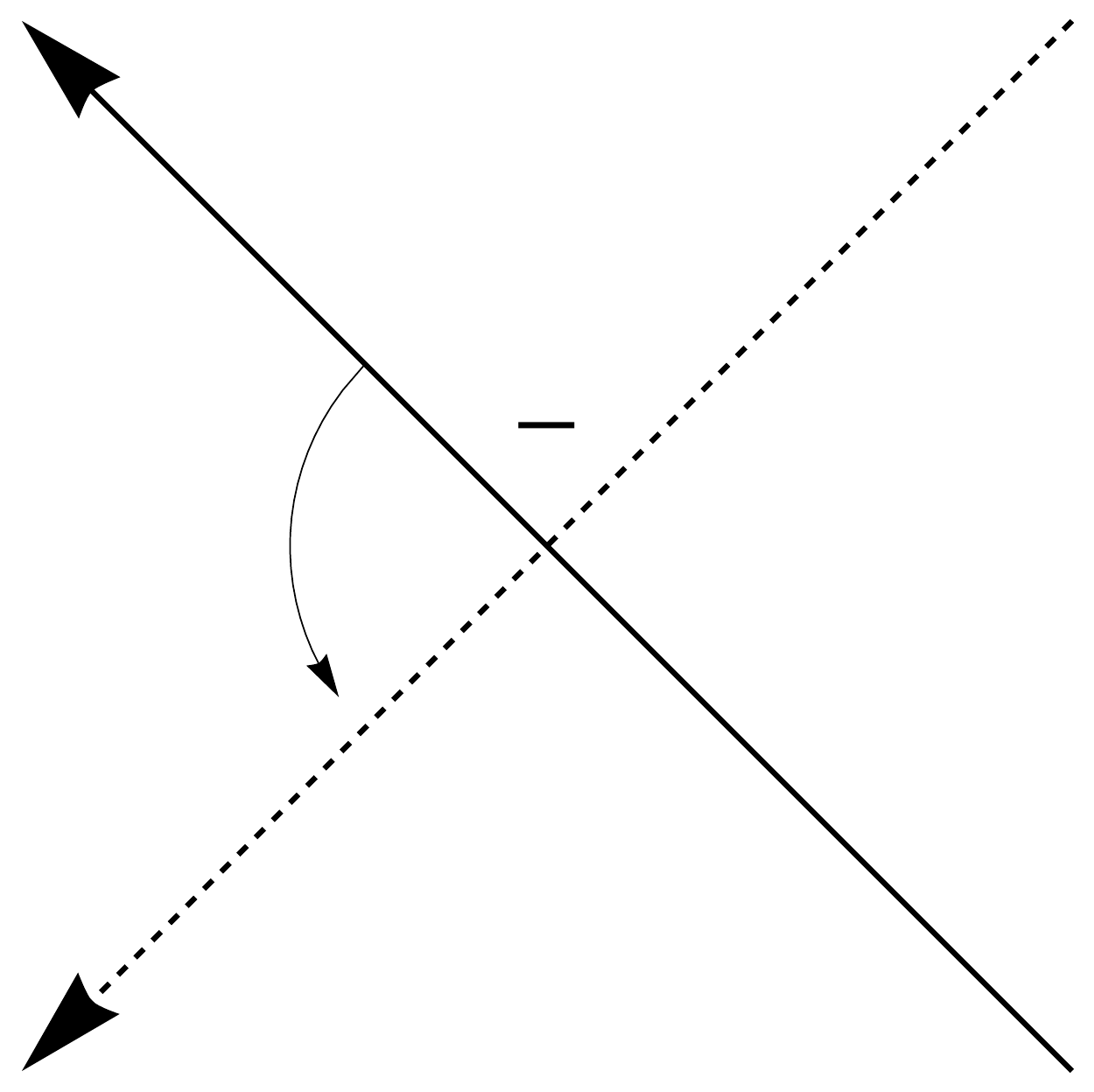,width=3cm,angle=-0} \\
\caption{Correspondence between chirality and signed crossing of loops: $a$ (black) and $b$ (dashed) type loops cross at a C site. Left: They anti-cross when the chirality of the black and white vertex sums to zero. Middle and right: They cross with a $\pm$ sign when the chirality is $\pm 1$.  }
\label{fig7}
\end{figure}

On the other hand, consider the winding loops of B-C type (called $a$-type, and represented by black lines in Fig.~\ref{fig7}) and winding loops  of C-A type (called $b$-type, and represented by dashed lines). If they cross, they cross in edges colored C. There are three situations, shown in Fig.~\ref{fig7}: when the chirality of the two vertices adds up to zero (left figure), there is no crossing at all. When the chirality is $+1$ (middle figure), the loops cross and the crossing has the $+$ sign. When the chirality is $-1$ (right figure), the sign is $-1$. Now if we consider all loops (winding and nonwinding) they cross (or anti-cross) in one third of all edges. A third of edges connects two vertices only once, so that summing over the C edges is equivalent to summing over all vertices. The total chirality is, therefore, equal to the number of signed crossings. Winding and nonwinding loops, on one hand, nonwinding and nonwinding loops, on the other  hand, have zero signed crossings; so that the total chirality is the number of signed crossings of the winding loops. The parity of the chirality and the parity of the number of crossings are the same invariant.

\subsection{Conservation of the norm modulo 4 for odd number of crossings}

Since the even sector is connected, we now restrict the configurations to those which have an odd number of crossings and define
\begin{eqnarray}
   \bm{a} . \bm{b} &=& a_x b_x+a_y b_y+a_zb_z \label{defproduct}\\
  &=& a_x b_x+a_y b_y+(a_x+a_y)(b_x+b_y) \\
  &=& 2a_xb_x+2a_yb_y+a_xb_y+a_yb_x. \label{odd}
\end{eqnarray}
$ \bm{a}. \bm{b}$ is necessarily an odd integer, because in Eq.~(\ref{odd}), $a_xb_y$ and $a_yb_x$ cannot be simultaneously odd or simultaneously even when the number of crossings, $a_xb_y-b_xa_y$ is odd: one is odd, one is even. With $ \bm{a}. \bm{b}$ odd, $ \bm{a}. \bm{b} \equiv 1 (\mbox{mod} 4)$ or $ \bm{a}. \bm{b} \equiv 3 (\mbox{mod} 4)$, which, following a standard convention, we note $-1 \equiv 3 (\mbox{mod} 4)$. So that $\bm{a}.\bm{b} \equiv \pm 1$(mod 4). Now, if we consider $ \bm{a}. \bm{c}$, we have
\begin{equation}
 \bm{a}. \bm{c} = - \bm{a}. \bm{b} -2(a_x^2+a_y^2+a_xa_y).
  \end{equation}
But $a_x^2+a_y^2+a_xa_y$ is odd, because $a_x$, or $a_y$, or both are odd. Because $ \bm{a}. \bm{b}$ is odd, we get a sign change and $ \bm{a}. \bm{c}\equiv  \bm{a}. \bm{b}$ (mod 4). Similarly,  $ \bm{a}. \bm{b}\equiv  \bm{b}. \bm{c}$ (mod 4). 
Hence we can define a symmetric version,
\begin{eqnarray}
I_2 =  \frac{1}{3} ( \bm{a}. \bm{b}+ \bm{b}. \bm{c}+ \bm{c}. \bm{a}) \hspace{.1cm} \mbox{mod} 4, \label{i2}
\end{eqnarray}
where the three terms are, in fact, equal, giving $I_2=\pm 1$ (recall $-1 \equiv 3$ (mod 4)). It can also be written by simple rearrangements,
\begin{eqnarray}
I_2 = - \frac{1}{6} ( |\bm{a}|^2+ |\bm{b}|^2+ |\bm{c}|^2) \hspace{.1cm} \mbox{mod} 4 = -n^2 \hspace{.1cm} \mbox{mod} 4. 
\end{eqnarray}
In this form, we see that $I_2$ is nothing but the norm squared of the total windings, modulo 4.

After a transformation along $a$, with a factor $k$ [Eq.~\ref{dl1}], we get,
\begin{eqnarray}
  I_2' &=& \frac{1}{3} ( \bm{a}'. \bm{b}'+ \bm{b}'. \bm{c}'+ \bm{c}'. \bm{a}') \hspace{.1cm} \mbox{mod} 4 \\
  &=& I_2 - k(k+\mbox{gcd}(a_x,a_y))\hat{ \bm{a}}^2 \hspace{.1cm} \mbox{mod} 4 .
\end{eqnarray}
We now show that the variation vanishes for all $k$.  First, since both
components $a_x$, $a_y$ cannot be simultaneously even in the odd sector, gcd$(a_x,a_y)$
is odd; and $k(k+\mbox{gcd}(a_x,a_y))$ is even. $\hat{ \bm{a}}^2$ results
from the definition of the scalar product, Eq.~(\ref{defproduct}):
$\hat{ \bm{a}}^2=2(a_x^2+a_y^2+a_xa_y)$ is even.  Therefore, $ k(k+\mbox{gcd}(a_x,a_y))\hat{ \bm{a}}^2 $ is a multiple of 4 and $I_2'=I_2$. The same argument applies for the other two transformations by symmetry. $I_2$ is therefore  a
conserved quantity and, since it takes two values, $I_2=\pm 1$, it defines two sectors, within the odd sector.

\subsection{Conservation of the sign for odd number of crossings}
\label{integers}

The third invariant is associated to the observation that the dynamics cannot reverse the sign: a given configuration $( \bm{a}, \bm{b}, \bm{c})$ and its \textit{antipodal} configuration $(- \bm{a}, -\bm{b}, -\bm{c})$, when it exists, belong to two different sectors. They cannot be distinguished by $I_2^a$ or $I_2$ which are both of even degree and have the same value for both configurations, so it must be proven by a different invariant. We define two quantities,
\begin{equation}
    I^{\pm} = \frac{1}{2}[I_6+I_5^{\pm} ] \mbox{mod} 4, \label{signinv}
 \end{equation}
where the sign $\pm$ refers to $I_2=\pm 1$, and
\begin{eqnarray}
    I_6 & = & a_xa_ya_zb_xb_yb_z+b_xb_yb_zc_xc_yc_z+c_xc_yc_za_xa_ya_z, \hspace{.7cm} \label{i6gen} \\ 
    I_5^{\pm} & = & (a_x \pm a_y)(a_y \pm a_z)(a_z \pm a_x)\chi  \nonumber \\
    &+&  (b_x \pm b_y)(b_y \pm b_z)(b_z \pm b_x)\chi \nonumber \\
    &+&  (c_x \pm c_y)(c_y \pm c_z)(c_z \pm c_x)\chi ,
    \label{i5gen}
 \end{eqnarray}
where $\chi = \frac{1}{3}( \bm{a}\times \bm{b}+ \bm{b} \times \bm{c}+
\bm{c} \times \bm{a})$ [Eq.~(\ref{chidef})] is odd. We will show that $I^{\pm}$ are indeed conserved by the dynamics in their respective sectors $I_2=\pm 1$ and allow to distinguish a configuration from its antipodal configuration.

In the odd sector, one and only one component in
each vector $\bm{a}$, $\bm{b}$ and $\bm{c}$ is even (see
Eq.~(\ref{pparity}) below). $I_6$, which contains the product of two such components, is therefore a multiple of 4, \textit{i.e.}
$\frac{1}{2} I_6 \equiv 0,2$ (mod 4). Similarly,  $I_5^{\pm}$ is an odd sum of products of four terms. In each product, one term and only one
is even, so that $\frac{1}{2}I_5^{\pm}$ is odd, hence $\frac{1}{2}I_5^{\pm}=\pm 1$(mod 4), and
therefore $I^{\pm}=\pm 1$.
It takes opposite signs for $( \bm{a}, \bm{b}, \bm{c})$ and $(-\bm{a},-\bm{b},-\bm{c})$. Indeed, the first term  $\frac{1}{2}I_6$ is unchanged, but the second term, which carries the sign, is reversed, so that $I^{\pm} \rightarrow -I^{\pm}$.
We have shown numerically that $I^{\pm}$ are conserved. They thus label the two ``antipodal'' sectors.
Here we give the explicit proof that $I^{+}$ is an invariant.
We first show that it is invariant in permutations, \textit{e.g.}
\begin{equation}
(\bm{a},\bm{b},\bm{c}) \rightarrow (-\bm{b},-\bm{a},-\bm{c}).
  \end{equation}
 $I_6$ is symmetric in the permutation of $\bm{a}, \bm{b}, \bm{c}$ and has even degree so the change of sign has no effect. The
term $(a_x \pm a_y)(a_y \pm a_z)(a_z \pm a_x)$ becomes $-(b_x \pm
b_y)(b_y \pm b_z)(b_z \pm b_x)$, but $\chi= \bm{a} \times \bm{b}
\rightarrow (-\bm{b}) \times (-\bm{a})=-\chi$ so the product is
unchanged.

We now consider the transformations $T_{i}(k)$ ($i=a,b,c$), with $k=\pm 1$, without
loss of generality. Since we have the invariance in permutations, we will always permute the configurations to an equivalent configuration identified by 
 a value of
$(\bm{a},\bm{b},\bm{c}) \mbox{mod} 2$. We choose, for instance,
 \begin{equation}
   (\bm{a},\bm{b},\bm{c}) \mbox{mod} 2= (1,0,1,0,1,1,1,1,0) \label{pparity}
\end{equation}
which is one of the six possible configurations for odd $\bm{a} \times \bm{b}$,
the other ones being obtained by permutations of the three vectors $(1,0,1)$, $(0,1,1)$, $(1,1,0)$. It has three even components $a_y$,
$b_x$ and $c_z$, so that $a_xb_y$ is odd, $a_yb_x$ even, and
the difference $\bm{a} \times \bm{b}$ is, indeed, odd.

For these configurations, $I^{\pm}$ simplifies and we give the simplified expression of $I^+$, \textit{i.e} when $I_2=+1$.

In order to simplify $I_6$, we use the fact that the product of an odd by an even number modulo 4 equals the even number: $(2p+1)\times 2n \equiv 2n (\mbox{mod} 4)$. We thus have,
\begin{equation}
\frac{1}{2} a_xa_ya_zb_xb_yb_z (\mbox{mod} 4) = \frac{1}{2} a_yb_x (\mbox{mod} 4), 
\end{equation}
because $a_xa_zb_yb_z$ is odd and $\frac{1}{2} a_yb_x$ even, so that
\begin{equation}
\frac{1}{2} I_6 (\mbox{mod} 4)=  \frac{1}{2} (a_yb_x+b_xc_z+c_za_y) (\mbox{mod} 4). 
\end{equation}
We show that the sum of the last two terms vanish. If $c_z$, which is even, is a multiple of 4, then $\frac{1}{2} c_z(b_x+a_y)\equiv 0 (\mbox{mod} 4) $ because $b_x+a_y$ is also even. If $c_z \equiv 2 (\mbox{mod} 4)$, we get $c_z=a_x+a_y+b_x+b_y \equiv 2 (\mbox{mod} 4)$.
Consider $a_x+b_y$.  
By using the definition $I_2=\bm{a}.\bm{b} (\mbox{mod} 4)$ and Eq.~(\ref{odd}), we get $I_2=a_xb_y (\mbox{mod} 4)=1$.
Since $a_xb_y=(2n+1)(2p+1)=1+2(n+p) (\mbox{mod} 4)$, we must have $n+p$ even. As a consequence, $a_x+b_y=2+2(n+p) \equiv 2 (\mbox{mod} 4)$, hence $b_x+a_y \equiv 0 (\mbox{mod} 4)$. Therefore, in this case too, $\frac{1}{2} c_z(b_x+a_y)\equiv 0 (\mbox{mod} 4) $ and 
\begin{equation}
\frac{1}{2} I_6 (\mbox{mod} 4)=  \frac{1}{2} a_yb_x (\mbox{mod} 4).
\end{equation}
We now consider $I_5^+$ [Eq.~(\ref{i5gen})], and replace $(a_x+a_y)$ by $-a_z$ etc.,
\begin{equation}
 I_5^+=- (a_xa_ya_z+b_xb_yb_z+c_xc_yc_z) \chi.
  \end{equation}
We note that,
\begin{equation} 3a_xa_ya_z=a_x^3+a_y^3+a_z^3. \label{eqsimple}\end{equation}
Since $(2n+1)^3 \equiv 2n+1 (\mbox{mod} 8)$ and $(2n)^3 \equiv 0 (\mbox{mod} 8)$, and since only
 $a_y$ is even, Eq.~(\ref{eqsimple}) reduces, modulo 8, to
\begin{equation}
 a_x^3+a_z^3 (\mbox{mod} 8)=a_x+a_z (\mbox{mod} 8)=-a_y (\mbox{mod} 8). 
\end{equation}
Now since $3\times 2n \equiv -2n (\mbox{mod} 8)$, we get $a_xa_ya_z (\mbox{mod} 8) \equiv a_y (\mbox{mod} 8)$. We thus obtain,
\begin{equation}
  \frac{1}{2} I_5^+ (\mbox{mod} 4)  = -\frac{1}{2}(a_y +b_x +c_z)\chi   (\mbox{mod} 4). \label{i00}
\end{equation}
Consider $\chi (\mbox{mod} 4)=a_xb_y-a_yb_x (\mbox{mod} 4)=I_2=+1$ for the current value of the parity. Note that $\chi$ is globally \textit{not} conserved modulo 4 (only modulo 2) because it changes sign in permutations. We can replace $\chi \equiv 1 (\mbox{mod} 4)$ and
combine the two equations,
\begin{eqnarray}
  I^+ 
  &=& \frac{1}{2}( a_yb_x-(a_y+b_x+c_z) )\mbox{mod} 4, \label{p1inv}
 \end{eqnarray}
which can be used provided that the configuration has the parity given by Eq.~(\ref{pparity}).
Since, however, it breaks the symmetry between the three types of loops, we have to consider the three types of transformations in turn.

\subsubsection{Invariance in the insertion of $a$-type loops}

The transformation $T_a(k=\pm1)$  is combined
with a permutation, $P_a.T_a(k=\pm 1)$,  in order to conserve the correct parity pattern of Eq.~(\ref{pparity}). Applying this transformation to a general configuration $(\bm{a},\bm{b},\bm{c})$, we obtain 
\begin{equation}
(\bm{a}',\bm{b}',\bm{c}')=(\bm{a}-2(g+k)\bm{\hat{a}},\bm{b}+(g+k)\bm{\hat{a}},\bm{c}+(g+k)\bm{\hat{a}}), \label{t1}
\end{equation}
where $g=\mbox{gcd}(a_x,a_y)$.  We see, indeed, that
\begin{equation} (\bm{a}',\bm{b}',\bm{c}') \mbox{mod} 2=(\bm{a},\bm{b},\bm{c}) \mbox{mod} 2,
\end{equation}
because $g$ is odd in the odd sector and $k=\pm 1$.
By using the transformation Eq.~(\ref{t1}) and the form [Eq.~(\ref{p1inv})] of the invariant, we obtain,
\begin{eqnarray}
  (I^+)_a'&=& \frac{1}{2}(a_y'b_x'-(a_y'+b_x'+c_z')) \mbox{mod} 4 \\ &=&
  I^+ + \frac{1}{2}(g+k)\hat{a}_y (g\hat{a}_x+3) \mbox{mod} 4,  
\end{eqnarray}
after removing two multiples of 4. Now, $\frac{1}{2}(g+k)$ is an integer, $\hat{a}_y$ is even according to Eq.~(\ref{pparity}), $g\hat{a}_x$ is odd (both are odd), so that $(g\hat{a}_x+3)$ is even and the product is a multiple of 4. Therefore $(I^+)_a'=I^+$.

\subsubsection{Invariance in the insertion of $b$-type loops}

Consider the action of the parity-conserving transformation $P_b.T_b(k=\pm 1)$,
\begin{equation}
(\bm{a}',\bm{b}',\bm{c}')=(\bm{a}+(g+k)\bm{\hat{b}},\bm{b}-2(g+k)\bm{\hat{b}},\bm{c}+(g+k)\bm{\hat{b}}), \label{tb}
\end{equation}
where $g=\mbox{gcd}(b_x,b_y)$. 
The invariant is similar,
\begin{equation}
  (I^+)_b'=
  I^+ + \frac{1}{2}(g+k) \hat{b}_x( g \hat{b}_y+3) \mbox{mod} 4,  
\end{equation}
where $(g+k)$, $\hat{b}_x$ and $(g\hat{b}_y+3)$ are all even. The variation therefore vanishes, $(I^+)_b'=I^+$.

\subsubsection{Invariance in the insertion of $c$-type loops}

Last, consider similarly the action of $P_c.T_c(k=\pm 1)$,
\begin{equation}
(\bm{a}',\bm{b}',\bm{c}')=(\bm{a}+(g+k)\bm{\hat{c}},\bm{b}+(g+k)\bm{\hat{c}},\bm{c}-2(g+k)\bm{\hat{c}}), \label{tc}
\end{equation}
where $g=\mbox{gcd}(c_x,c_y)$.
The invariant becomes,
\begin{equation}
  (I^+)_c'=
  I^+ + \frac{1}{2}(g+k) ( a_y \hat{c}_x+ b_x \hat{c}_y  + (g+k) \hat{c}_x\hat{c}_y+ 3\hat{c}_z) \mbox{mod} 4.  \nonumber
\end{equation}
The terms in brackets are even, because $(g+k)$ is even, and $a_y$, $b_x$, and $c_z$ are the three even components. If $\frac{g+k}{2}$ is even, then the variation is a multiple of 4 and vanishes. Consider  $\frac{g+k}{2}$ odd, in the rest of the proof. By using, again, $(2p+1)\times 2n \equiv 2n (\mbox{mod} 4)$,  we get
\begin{eqnarray}
  (I^+)_c' &=&  I^+ + ( a_y \hat{c}_x +b_x \hat{c}_y + (g+k) \hat{c}_x\hat{c}_y+3\hat{c}_z) \mbox{mod} 4  \nonumber \\
  &=&  I^+ + (  a_y + b_x + (g+k) - \hat{c}_z) \mbox{mod} 4,  \nonumber
  \end{eqnarray}
where, in the second line, we have eliminated the odd factors multiplied by even numbers. Note that $g+k=4n+2 \equiv 2 (\mbox{mod} 4)$. Note also that $c_z=g \hat{c}_z=(4n+1)\hat{c}_z$ where $\hat{c}_z$ is an integer by definition, implying $\hat{c}_z \equiv c_z (\mbox{mod} 4)$. We now replace $\hat{c}_z$ by $c_z=a_x+a_y+b_x+b_y$,
\begin{eqnarray}
  (I^+)_c'
  &=&  I^+ + ( 2 - a_x-b_y) \mbox{mod} 4.  \nonumber 
   \end{eqnarray}
Since $a_x+b_y \equiv 2 (\mbox{mod} 4)$ (see above), we have 
therefore $(I^+)_c'=I^+$.

\subsection{Number of sectors from the invariants}

The three invariants $I_2^a$, $I_2$ and $I^{\pm}$ allow to label the
Kempe sectors. We have $I_2^a=0$ for the even sector which is
connected, and $(I_2^a,I_2,I^{\pm})=(1,\pm 1, \pm 1)$ for the odd
sectors. We thus have five sectors that define disconnected
configurations.

In table~\ref{table1}, we give examples of the winding numbers with smallest
norm $n^2$ [as defined by Eq.~(\ref{normdef})],  to show that each sector is, indeed, realized
(provided that $L$ is large enough). We note that, when $L$ is not a multiple of three, antipodal sectors do not always exist, but different sectors of $I^{\pm}$ do exist.
We also show the
labels of the five large sectors in Fig.~\ref{fig5}.

Since we have enlarged the number of winding configurations compared with the coloring problem, we have checked that all the values of the invariants are also realized in the coloring problem (see table~\ref{tableKT}).
\begin{table}
\begin{center}
\begin{tabular}{r|r|r|r|r}
$I_2^a$ & $I_2$  & $I^{\pm}$ &  $( \bm{a}, \bm{b}, \bm{c})$    & $n^2$ \\
  \hline
0 & - & - & 0 & 0 \\  
1 &  1 & $\pm $1 &  $\pm  (2,-1,-1,-1,-1,2,-1,2,-1)$ & 3 \\
1 &  -1 & $\pm $1 & $\pm  (3,-1,-2,-3,2,1,0,-1,1)$   & 5 \\
\end{tabular}

\begin{tabular}{r|r|r|r|r}
$I_2^a$ & $I_2$  & $I^{\pm}$ &  $( \bm{a}, \bm{b}, \bm{c})$ & $n^2$ \\
  \hline
0 & - & - & $(0,0,0,-1,-1,2,1,1,-2)$ & 2 \\
  1& 1 & $\pm$ 1 & $\pm (-3,2,1,2,1,-3,1,-3,2)$ &  7 \\
  1& -1 & $-1$ & $(-1,0,1,1,-1,0,0,1,-1)$ &  1 \\
1&  -1 & $1$  & $(-3,4,-1,5,-3,-2,-2,-1,3)$ &  13 
 \end{tabular}

\begin{tabular}{r|r|r|r|r}
$I_2^a$ & $I_2$  & $I^{\pm}$ &  $( \bm{a}, \bm{b}, \bm{c})$  & $n^2$ \\
  \hline
  0 & - & - & $(0,0,0,1,1,-2,-1,-1,2)$ & 2 \\
1& 1 & $\pm$ 1 & $\pm (-3,1,2,4,-1,-3,-1,0,1)$ &  7 \\
  1& -1 & $1$ & $(1,0,-1,-1,1,0,0,-1,1)$ &  1 \\
1& -1 & $-1$  & $(-1,1,0,3,2,-5,-2,-3,5)$ &  13 \\
 \end{tabular}

\end{center}
\caption{Examples of configurations of winding numbers $( \bm{a}, \bm{b}, \bm{c})$ with the smallest norm $n^2$, in the Kempe sectors labeled by $(I_2^a,I_2,I^{\pm})$, for $L \mbox{mod} 3=0,1,2$ (from top to bottom). One can check that $\bm{a}+\bm{b}+\bm{c}=0$, $a_x+a_y+a_z=0$, $\bm{a}-\bm{b} \equiv L$(mod 3) etc.}
\label{table1}
\end{table}

There are still many disconnected sectors 
labeled by the same invariants (see Figs.~\ref{fig4} and \ref{fig5}) and this point is studied in
section~\ref{steric}.

\subsection{Symmetries}

The two large sectors
labeled by $(1,1,\pm 1)$ have the same size, because they are the chiral image of each other in mirror plane symmetries $P$. For a given
configuration of winding numbers, there is always another
configuration obtained by exchanging two axis among the $x$, $y$ or $z$ axis. When $I_2=1$, we see that $I^+_5$ in Eq.~(\ref{i5gen}) is reversed in an exchange of axis (because $\chi$ takes a minus sign). Hence $I^+ \rightarrow -I^+$ and the two mirror configurations are in separate sectors. Note, indeed, that there are special states $(\bm{a},\bm{b},\bm{c})$ with $a_y=a_z$, $b_y=c_z$, $b_z=c_y$, $b_x=c_x$ (see the example in the top table of Table~\ref{table1}),  which a mirror plane $yz$ transforms effectively according to $(\bm{a},\bm{b},\bm{c}) \rightarrow (\bm{a},\bm{c},\bm{b})$. An allowed permutation  gives in turn $(\bm{a},\bm{c},\bm{b}) \rightarrow (-\bm{a},-\bm{b},-\bm{c})$ which is the opposite of the original state. The antipodal transformation $T$, therefore, results in a lattice symmetry operation when $I_2=1$.  As a consequence, the
two sectors must have the same size.
This symmetry implies an equal number of homotopy classes in the two sectors (see the degeneracy in Fig.~\ref{fig5}), and also an equal number of color configurations (see the multiplicity in table~\ref{tableKT}).  
Since a configuration in one of these sectors and its mirror image are not connected, the dynamical evolution remains chiral.

The two other sectors $(1,-1,\pm 1)$ are nondegenerate, however.
When $I_2=-1$, a
configuration and its mirror image are in the same sector: $I_5^- \rightarrow I_5^-$ [Eq.~(\ref{i5gen})] ($\chi$ cancels the sign of
say $a_x-a_y$), hence $I^-$ is invariant in mirror planes.
In this case, the antipodal transformation is not a symmetry. Note that,
for a given $( \bm{a}, \bm{b}, \bm{c})$, $( -\bm{a}, -\bm{b},
-\bm{c})$ may violate the constraints (see
Eqs.~(\ref{eq1})-(\ref{eq3})), so that the asymmetry of the
constraints implies that of the two sectors $(1,-1,\pm 1)$.

The dynamics conserves $I^+=\pm 1$ which distinguishes two degenerate chiral sectors, and $I^-=\pm 1$ that defines two antipodal achiral nondegenerate sectors.

\section{Steric constraints, dilute limit}
\label{steric}

We argue that all the remaining disconnected sectors (see Fig.~\ref{fig5}), which do share the same invariants $(I_2^a,I_2,I^{\pm})$, cannot be distinguished by any other invariant. The remaining obstruction is not a property of the transformations, but is of steric origin.

\begin{figure}[h] \vspace{0cm} \center 
\psfig{file=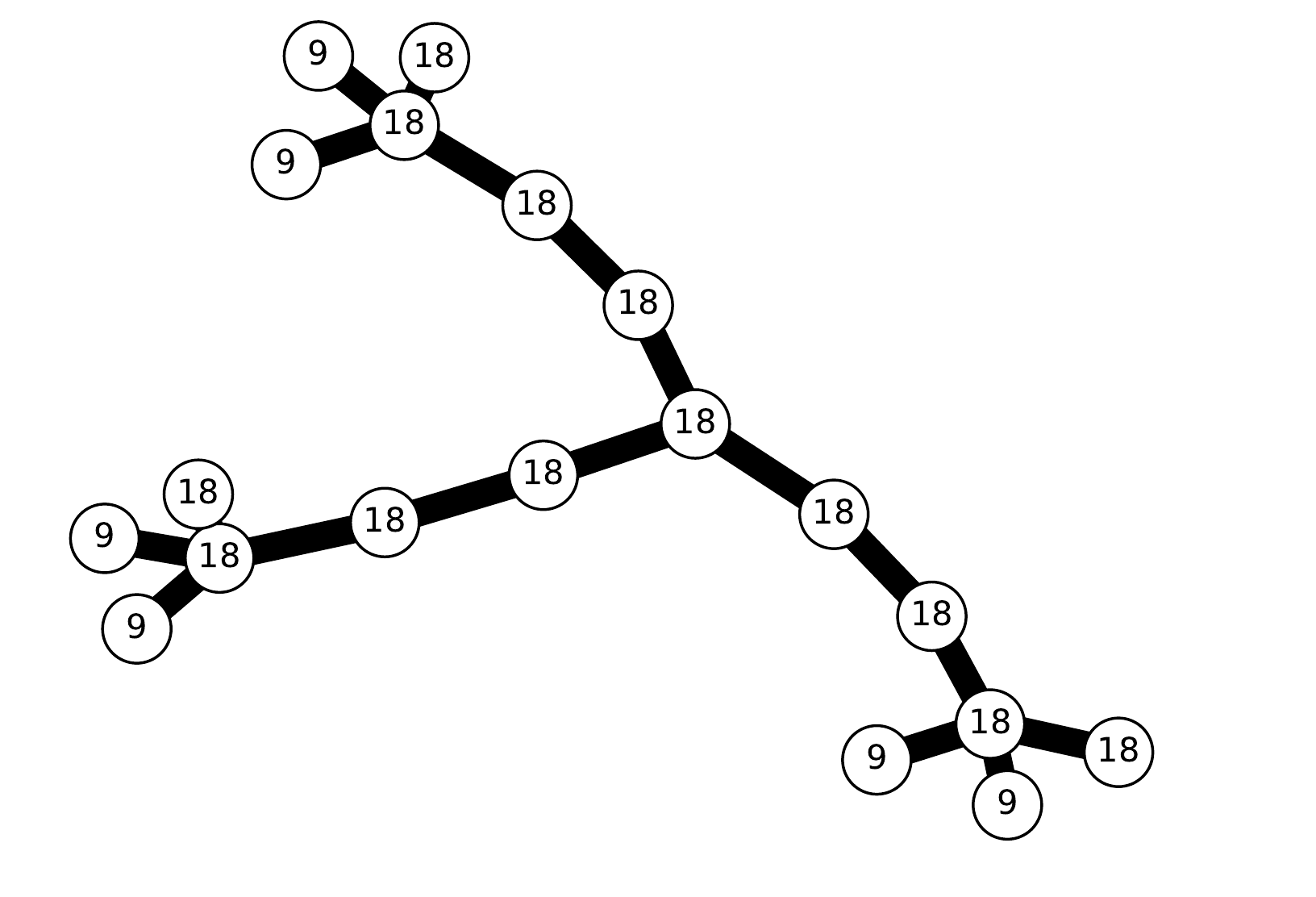,width=8.0cm,angle=-0} 
\caption{Example of a Kempe sector with $(I_2^a,I_2,I^{\pm})=(1,-1,1)$ for $L=18$, represented as a graph: nodes (winding number configurations $( \bm{a}, \bm{b}, \bm{c})$, up to permutations) are connected by the dynamics. Indicated is the minimal size $L$ for the configuration to fulfill the constraints, thus showing that the paths between $L=9$ configurations are cut, giving $1 \rightarrow 6$ isolated sectors by steric obstruction.}
\label{fig80}
\end{figure}

We show this by allowing larger sizes at fixed and small winding
numbers. As earlier, we construct the Kempe sectors for a given size
$L$. They are each characterized by a set of winding numbers up to
$L$. The same set of winding numbers can be considered at larger sizes
$L_1 \geq L$ (provided that $L_1 \equiv L (\mbox{mod} 3)$). We can now
redo the construction of the sectors starting with the same winding
numbers. In the dynamics, larger winding numbers are generated and we
test whether they open new paths between the original winding
configurations. In Fig.~\ref{fig80}, we give an example of the graph
of a Kempe sector. The nodes are winding loop configurations $(
\bm{a}, \bm{b}, \bm{c})$ (up to permutations), connected by edges when
a dynamical transformation between them exists. The number indicated
in each node is the minimal size $L$ for the winding numbers to
fulfill the size constraints. We see then clearly that the winding
numbers existing for $L=9$, for instance, get reconnected through
winding number configurations that exist only for $L_1 \geq L$. The
path between them implies configurations with higher winding numbers.

\begin{figure}[h] \center \hspace{1cm}
\psfig{file=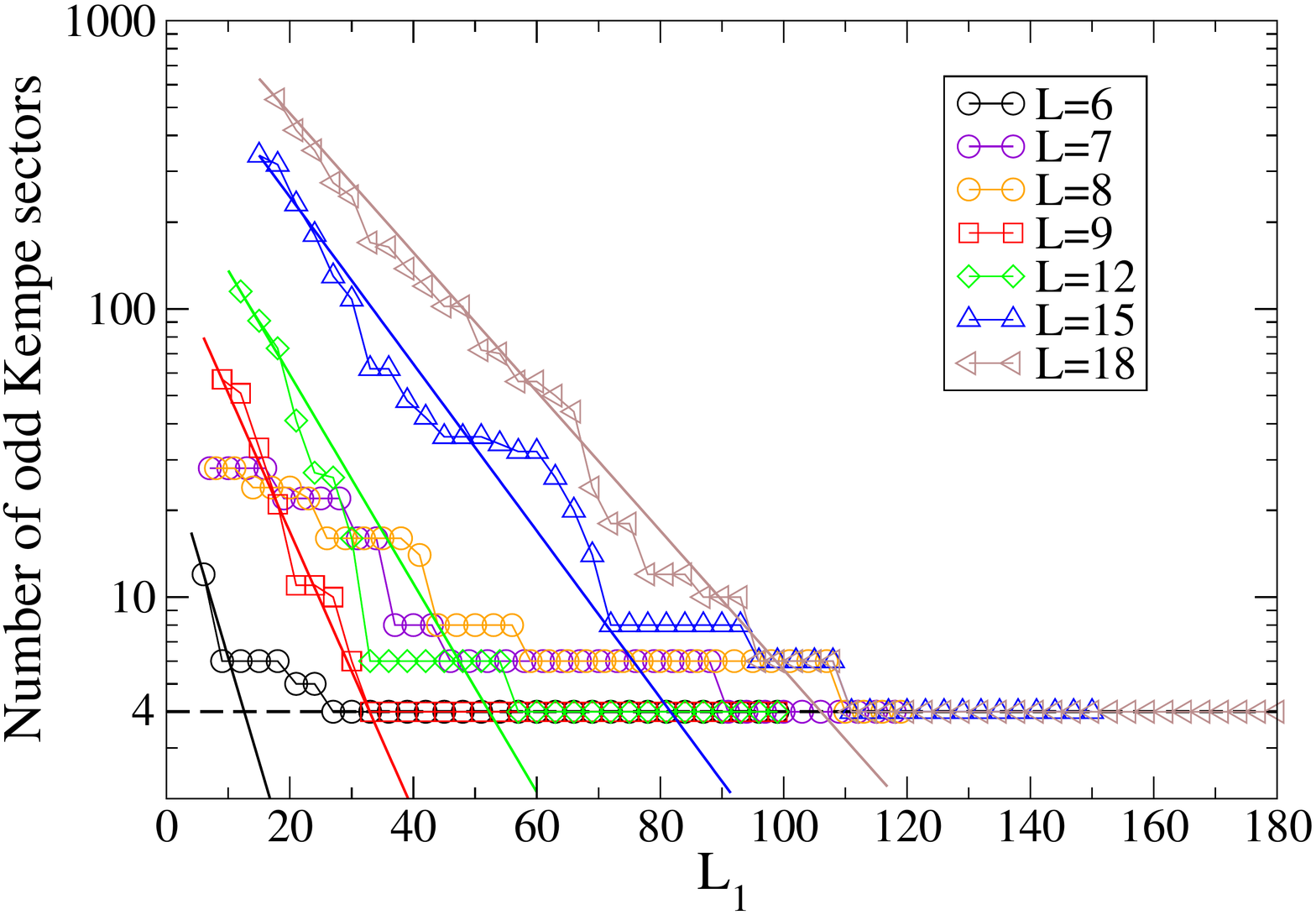,width=11.0cm,angle=-0} 
\caption{Multiscale reconnection of the odd Kempe sectors at size $L$ when winding numbers up to $L_1$ are allowed. By increasing $L_1$, the number of sectors decrease, showing reconnections, down to the four (dashed line) odd sectors characterized by the invariants. Decays as $\sim \exp(-L_1/L)$ are shown for comparison (solid lines).}
\label{fig8}
\end{figure}
\begin{figure}[h] \center
  \psfig{file=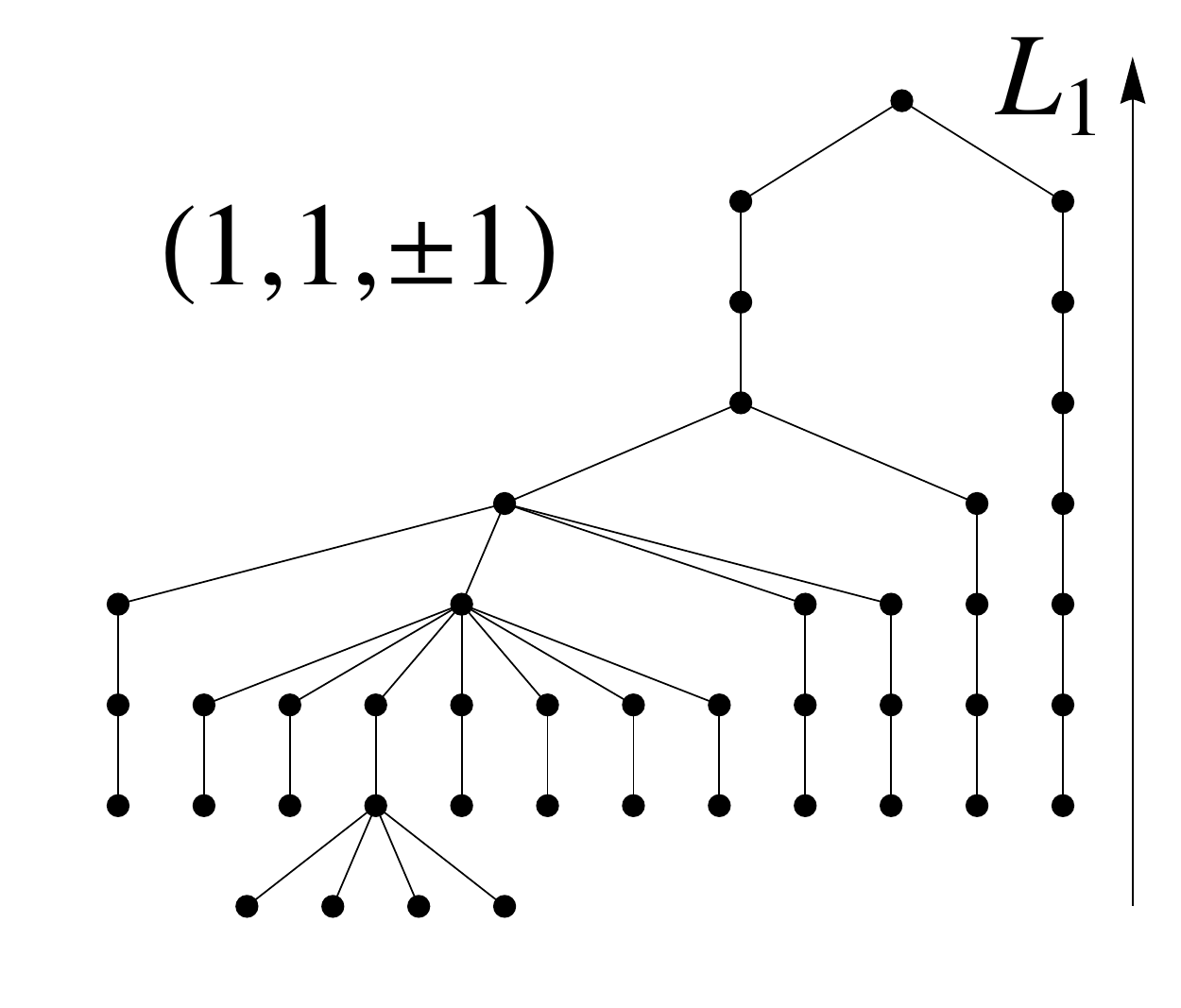,width=4.9cm,angle=-0}
  \psfig{file=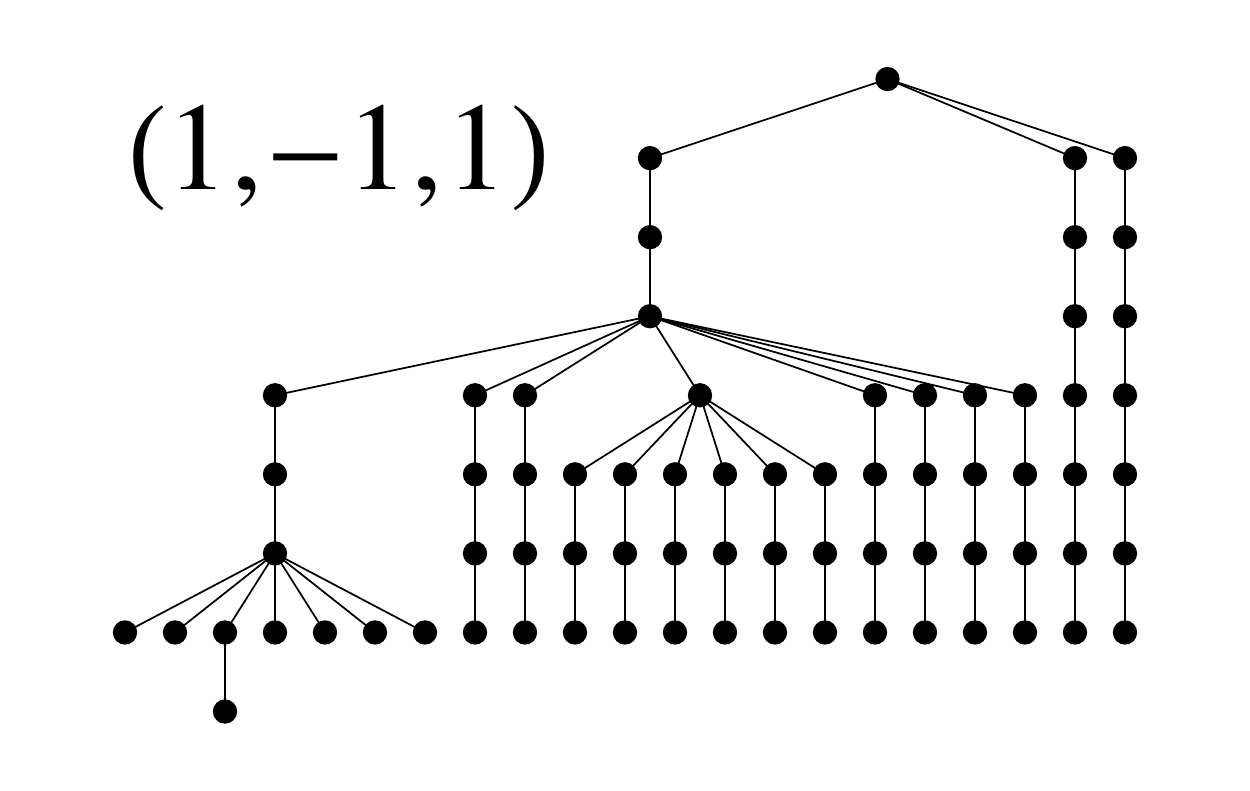,width=4.9cm,angle=-0}
  \psfig{file=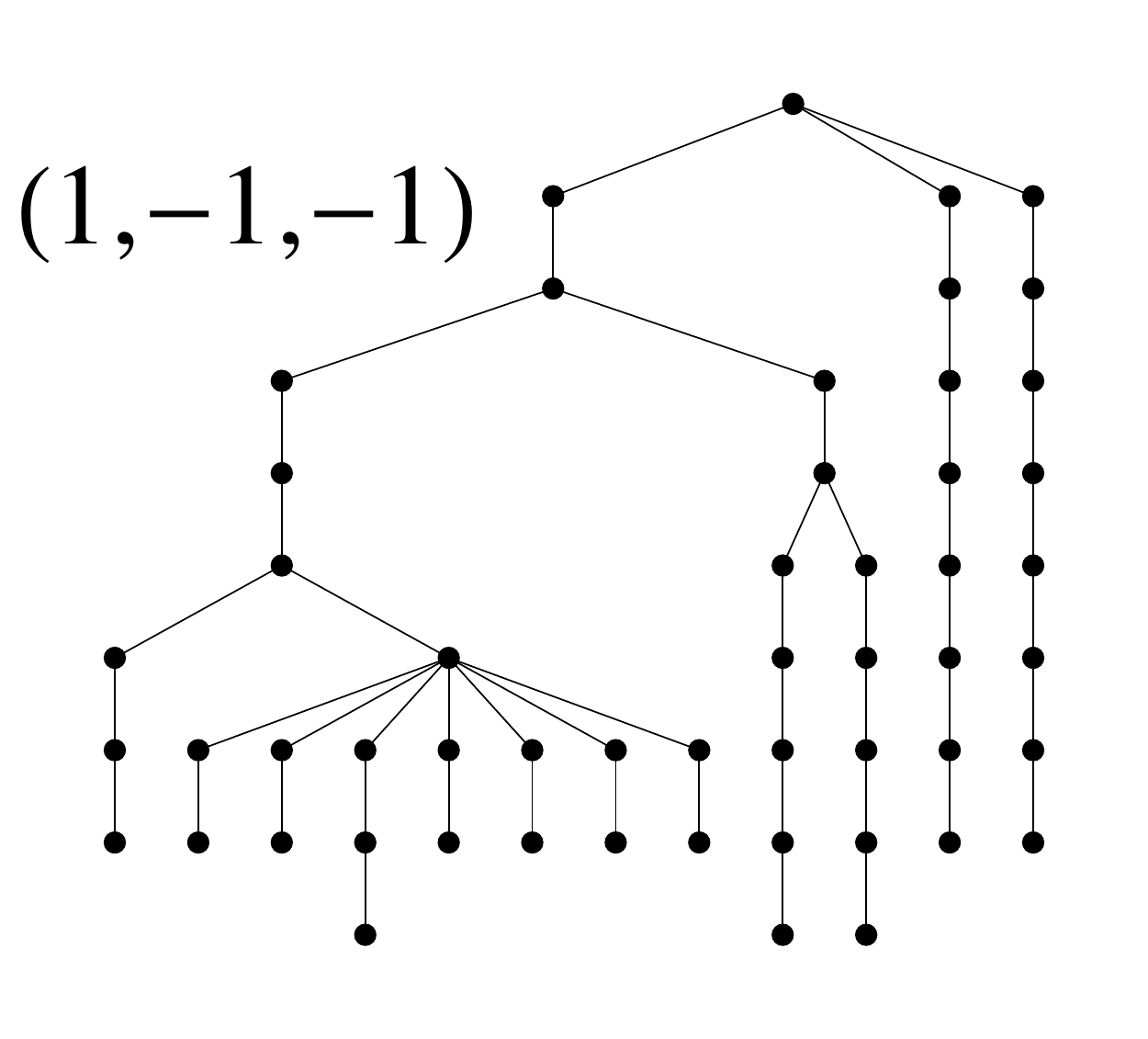,width=4.9cm,angle=-0} 
 \caption{Configuration space of low-winding numbers as four hierarchical trees (one for each invariant $(1,\pm 1,\pm 1)$, two are identical by symmetry). Each point represents a Kempe sector. The level $n \geq 0$, from bottom up corresponds to the system size, $L_1=6+3n$, \textit{i.e.} higher winding numbers allowed. Only the lowest two levels $n=0,1$ include \textit{all} the sectors. For $n\geq 2$, additional sectors are not shown.}
\label{fig81}
\end{figure}

In Fig.~\ref{fig8}, we see that, as the size $L_1$ increases, the
number of sectors constructed for size $L$ decrease, so that new
connections indeed appear. Moreover, in all cases considered, all
isolated odd sectors merge into four sectors (characterized by the
invariants of section~\ref{invariants}) for large enough $L_1$ (dashed
line), \textit{i.e.} when the system is diluted, $|\bm{a}|, |\bm{b}|,
|\bm{c}| \ll L_1$. The fact that the process converges to four means
that there are no other invariants (it cannot go lower than 4 because
of the distinct invariants). There exist paths between the winding
loop configurations, provided that the size is large enough. In other
words, the disconnection at size $L$, beyond that due to the
invariants, is a steric obstruction. This is illustrated in
Fig.~\ref{fig81} where all the odd Kempe sectors at $L=6,9$ are
represented by points in the first two bottom levels. The level $n$
corresponds to the size $L_1=6+3n$ ($n \geq 0$).  The lower levels get
reconnected at higher levels (through higher winding numbers). Note
that for $n \geq 2$ not all the sectors are shown.

While this occurs for the winding number configurations, the
reconnection could take place through unphysical configurations (see
section~\ref{numconstruction}) and may be unphysical for the coloring
problem. As noted earlier, the configurations in the dilute limit are
not expected to be constrained by the discreteness of the lattice, so
that we expect all sets of winding numbers in this limit.

The restoration of ergodicity (equilibration among the
sterically-obstructed sectors) is described in Fig.~\ref{fig8} by an
exponential decrease $\sim \exp(-L_1/\xi)$, where $\xi=L$ fits
approximately (solid lines). It means that the path between any two
loop configurations with low winding numbers implies configurations
with higher winding numbers of order a few $\xi$ (which we could think
of as higher energy states, if an energy were to be defined). Since
there is a distribution of higher winding numbers, the equilibration
is thus a multiscale process. The image is that of a space of
configurations consisting of hierarchical trees (Fig.~\ref{fig81}). A
physical model of equilibration of configurations with low winding
numbers may thus involve the ultrametric distance between sectors: the
dynamics is expected to be slow \cite{Rammal}.

\begin{figure}[ht] \center  \hspace{1cm}
\psfig{file=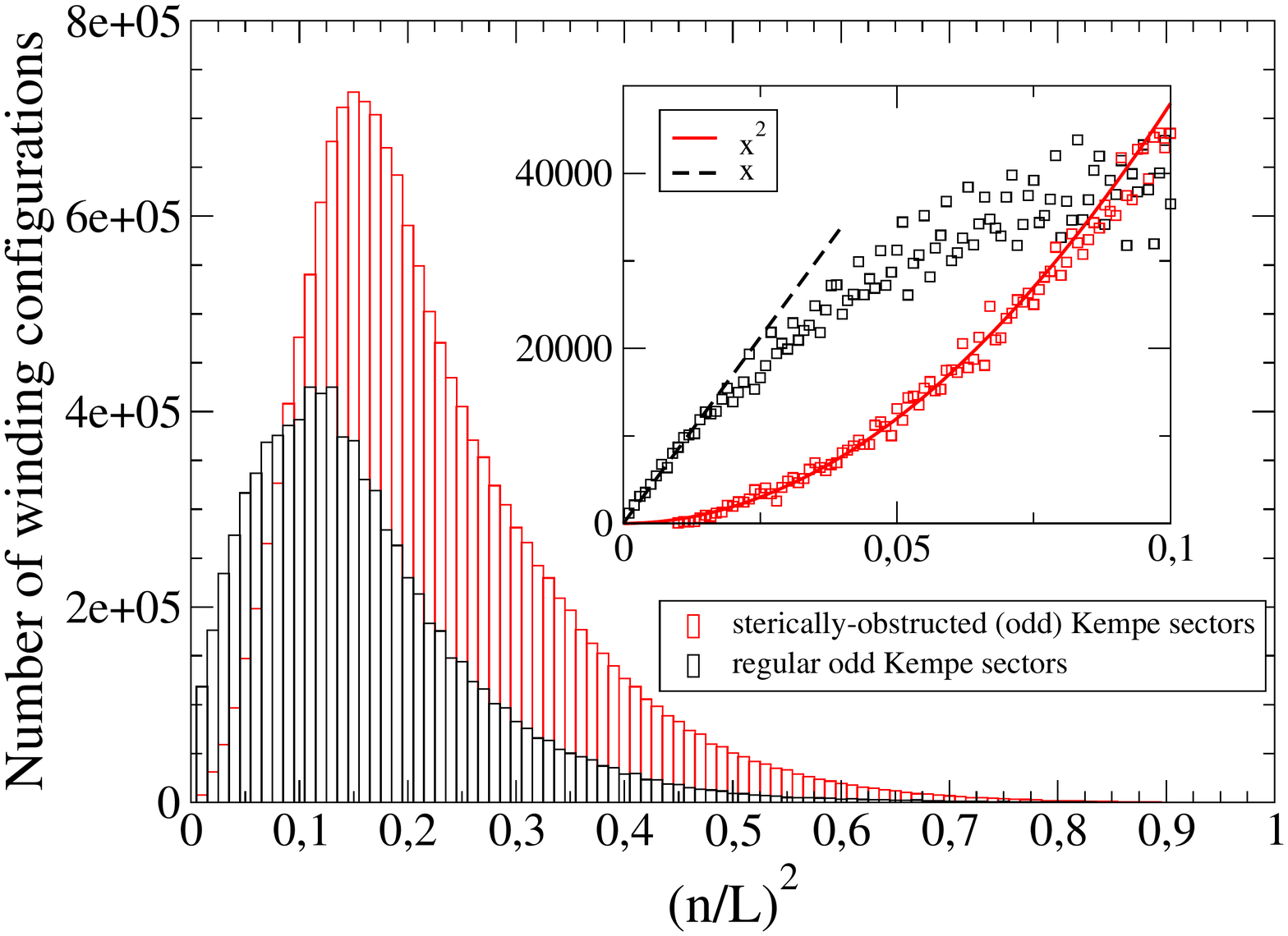,width=11.0cm,angle=-0} 
\caption{Probability distribution of norms, $n^2$ in winding loop configurations, for regular odd sectors and sterically-obstructed (odd) Kempe sectors ($L=150$). Small norms (see inset) are absent from the latter: linear (dashed line) and quadratic (solid line) fits are added. }
\label{fignorms}
\end{figure}

In order to confirm the connectedness of the space of configurations
in the dilute limit for larger $L$, we can study the distribution of
winding numbers. In Fig.~\ref{fignorms}, we give the probability
distribution function of the norm squared, $n^2=\frac{1}{6} (
|\bm{a}|^2+ |\bm{b}|^2+ |\bm{c}|^2)$, over all configurations for
$L=150$, by distinguishing configurations of regular or
sterically-obstructed sectors. Overall, it is apparent that a
randomly-chosen odd configuration has more chance to belong to a
sterically-obstructed sector. The probability for a configuration to
be odd is 0.37 (0.63 to be even), and within the odd sector, the
probability is 0.65 to belong to a sterically-obstructed odd sector
(0.35 to a regular odd sector).  Sterically-obstructed sectors have
all norms above a small size-dependent threshold.  In the limit of
small $(\frac{n}{L})^2 \ll 1$ (inset of Fig.~\ref{fignorms}), the
number of such sectors is vanishingly small, in $\sim (\frac{n}{L})^4$
(solid line), whereas the number of regular sectors varies like
$(\frac{n}{L})^2$ (dashed line). A randomly-chosen configuration
therefore belongs to a regular sector with a probability $1-
\alpha(\frac{n}{L})^2 \rightarrow 1$ in the limit $n \ll L$, and is
hence connected to all other configurations with the same label. This
confirms the existence of a proper ergodic dilute limit where the only
obstruction to ergodicity is that described by the invariants. On the
other hand, the probability to belong to a sterically-obstructed
sector increases steadily as $n$ increases, in agreement with the
picture that there is less and less space when $n \rightarrow L$ and
that steric obstruction is more and more prominent.

\section{Conclusion}

We have obtained a classification of the Kempe sectors on the regular
hexagonal lattice for generic sizes, based on the loop winding
numbers. We have found a set of five sectors labeled by three
invariant parities that are polynomials of the winding numbers. The
first invariant has a simple geometrical interpretation, this is the
parity of the number of crossings of the winding loops. When this
number is even, the sector is connected. When it is odd, it splits
into four sectors, two achiral and two chiral.  The two chiral sectors
have the same number of classes (degenerate) as they are image of each
other by lattice mirror symmetries.  Thus, there are configurations
that break mirror symmetries, that the dynamics cannot restore
(contrary to spontaneous symmetry breaking).  The two achiral sectors
do not break the mirror symmetry and are nondegenerate. They are
``antipodal'', a symmetry that is broken by the ``box'' constraints.
We assume that the invariants may be viewed geometrically as linking
numbers of flux lines on the lattice, which is embedded (or immersed)
into a 3d manifold.

At fixed $L$, there are many additional undistinguished sectors that
share the same invariants. We have shown that they would be connected
to the four odd sectors if there was enough space to accommodate
higher winding numbers. For these additional sectors, the problem is
therefore a steric obstruction, not a conservation law. We have thus
argued that the set of three invariants is complete.  This allowed to
describe the configuration space by distinguishing dilute from more
dense configurations.  While dense configurations can be
sterically-obstructed, a randomly-chosen configuration in the dilute
limit $|\bm{a}|, |\bm{b}|, |\bm{c}| \ll L$ (where space is available)
belongs to one of the five sectors and can be connected to any other
with the same label, with no further obstruction. Moreover, the
reconnection dynamics takes place on hierarchical trees and involves a
multiscale distribution of intermediate winding numbers. For this
reason, the equilibration dynamics of the low-flux lines, up to the
conservation laws, is expected to be ``slow''. This gives some
perspective as to how an entanglement of flux lines could lead to slow
dynamics \cite{Nelson}: here it results from the need of a
distribution of higher-flux intermediate configurations.

\section*{Acknowledgements}
O.C would like to thank P. Dehornoy and R. Bacher for their helpful guidance.

\begin{appendix}
\section{Sector with even number of line crossings is connected}
\label{connected}

We consider here the winding number configurations $( \bm{a}, \bm{b}, \bm{c})$ with $\bm{a} \times \bm{b}$ \textit{even} and show that they are all connected by the dynamics, Eqs.~(\ref{dl1})-(\ref{dl3}). We restrict the proof for $L$ a multiple of three.

 Start from the ``0-sector'' defined by
 $\bm{a}=\bm{b}=\bm{c}=(0,0,0)$, which exists only when $L$ is a
 multiple of three. The sector
 is even since $\bm{a} \times \bm{b}=0$. It will be our target state. Create a pair of $a$-type winding
 loops,
\begin{eqnarray}
\bm{a} &=& 2 \bm{m} \\
\bm{b} &=& -\bm{m} \\
\bm{c} &=& -\bm{m},
\end{eqnarray}
where $\bm{m}$ is an arbitrary vector. A second transformation adds a pair of $b$-type loops with the same $\bm{m}$,
\begin{eqnarray}
\bm{a} &=& 3 \bm{m}\\
\bm{b} &=& -3 \bm{m} \\
\bm{c} &=&  0.
\end{eqnarray}
Now we can add a pair of $c$-type winding loops with a different $\bm{m}_1$,
\begin{eqnarray}
\bm{a} &=& 3 \bm{m} - \bm{m}_1 \label{lasteq00}\\
\bm{b} &=& -3 \bm{m} -\bm{m}_1 \label{lasteq0} \\
\bm{c} &=& 2 \bm{m}_1, \label{lasteq}
\end{eqnarray}
and $\bm{a} \times \bm{b} =6 \bm{m}_1 \times \bm{m}$ is even (since the parity is invariant).

Conversely, consider $L$ a multiple of 3, and any $(\bm{a},\bm{b},\bm{c})$ representing  an even configuration.
When $\bm{a} \times \bm{b}=a_xb_y-a_yb_x$ is even, we
have to inspect 10 out of 16 parity configurations for the four independent integers $(a_x,a_y,b_x,b_y)$.

(i) they are all
odd. Then $\bm{c}=-\bm{a}-\bm{b}$ has three even components. Define
$\bm{m}_1=\bm{c}/2$, it is an integer vector, so is
$\bm{a}+\bm{c}/2=(\bm{a}-\bm{b})/2$. When $L$ is a multiple of 3,
the three components of $\bm{b}-\bm{a}$ are multiples of 3, according
to Eq.~(\ref{eq3}). We can therefore define an integer vector $\bm{m}$
such that $3\bm{m}=\bm{a}+\bm{c}/2$.
Now the vectors
$(\bm{a},\bm{b},\bm{c})$ have the form of
Eqs.~(\ref{lasteq00})-(\ref{lasteq}). Inverting all the transformations above,
 transforms this sector onto the
0-sector. 

(ii) they are all even. Again we can define
$\bm{m}_1=\bm{c}/2$, and the same argument applies. 

(iii) $a_x$
even, $b_y$ odd, and $a_y$ even, $b_x$ odd. Then
$\bm{c}=-\bm{a}-\bm{b}$ is odd. But $\bm{a}$ is even, so define
$\bm{m}_1=\bm{a}/2$ and the same argument applies again up to a
permutation. 

(iv) $a_x$ even, $b_y$ odd, and $a_y$ odd, $b_x$
even. Then $c_x=-a_x-b_x$ is even, and $c_y=-a_y-b_y$ is even. Define
$\bm{m}_1=\bm{c}/2$ etc. The other cases are similar.

Therefore any set of winding numbers $(\bm{a},\bm{b},\bm{c})$ such that $\bm{a} \times \bm{b}$ is even can be cast into the form of Eqs.~(\ref{lasteq00})-(\ref{lasteq}) with the definition of two vectors $\bm{m}, \bm{m}_1$ which by three transformations can be transformed onto the ``0-sector''. Therefore, any even configuration is connected to any other even configuration.

\end{appendix}

\nolinenumbers

\begin{thebibliography}{99}
\bibitem{Stewart} I. Stewart, Symmetry: A Very Short Introduction, Oxford University Press, \doi{10.1093/actrade/9780199651986.001.0001}.
\bibitem{Huse} D. A. Huse and A. D. Rutenberg, Phys. Rev. B \textbf{45}, 7536 (1992), see reference [13] therein, \doi{10.1103/PhysRevB.84.020413}.
\bibitem{Moore} C.~Moore and M.~E.~J.~Newman, J. Stat. Phys. 99, 629-660~(2000), \doi{10.1023/A:1018638624854}.
\bibitem{Moharinv} B. Mohar and J. Salas,  J. Phys. A: Math. Theor. 42 (2009) 225204, \doi{10.1088/1751-8113/42/22/225204}.
\bibitem{Mohar} B. Mohar and J. Salas, J. Stat. Mech. P05016 (2010), \doi{10.1088/1742-5468/2010/05/P05016}.
\bibitem{Normand} B. Normand, Phys. Rev. B \textbf{83}, 064413 (2011), \doi{10.1103/PhysRevB.83.064413}.
\bibitem{dimer} O. Sikora, N. Shannon, F. Pollmann, K. Penc, and P. Fulde, Phys. Rev. B \textbf{84}, 115129 (2011),  \doi{10.1103/PhysRevB.84.115129}.
\bibitem{Freedman} M. Freedman, M. B. Hastings, C. Nayak, and X.-L. Qi, Phys. Rev. B \textbf{84}, 245119 (2011), \doi{10.1103/PhysRevB.84.245119}.
\bibitem{Cepas} O. C\'epas and A. Ralko, Phys. Rev. B \textbf{84}, 020413 (2011), 		      \doi{10.1103/PhysRevB.84.020413}.
\bibitem{Baxter}  R.~J.~Baxter, J. Math. Phys. \textbf{11}, 784 (1970), \doi{10.1063/1.1665210}.
\bibitem{kagome} The sites of the kagome lattice are the centers of the edges of the hexagonal lattice.
\bibitem{Chandra} J.~T.~Chalker, P.~C.~W.~Holdsworth, and E.~F.~Shender, Phys. Rev. Lett. \textbf{68}, 855  (1992), \doi{10.1063/1.353624}. P.~Chandra, P.~Coleman, I.~Ritchey, J. Phys. I France \textbf{3}, 591 (1993), \doi{10.1051/jp1:1993104}.  
\bibitem{Chakraborty} B. Chakraborty, D. Das, and J. Kondev, Eur. Phys. J. E \textbf{9}, 227 (2002), \doi{10.1140/epje/i2002-10071-7}.
\bibitem{Castelnovo} C. Castelnovo, P. Pujol, and C. Chamon, Phys. Rev. B \textbf{69}, 104529 (2004), \doi{10.1103/PhysRevB.69.104529}.
\bibitem{Cepasglasses} O.~C\'epas, Phys. Rev. B \textbf{90}, 064404 (2014),  \doi{10.1103/PhysRevB.90.064404}.
\bibitem{Belcastro} S.-M.~Belcastro and R. Haas, Discrete Mathematics, \textbf{325}, 77 (2014), \doi{10.1016/j.disc.2014.02.014}.
\bibitem{oddeven} O. C\'epas, Phys. Rev. B \textbf{95}, 064405 (2017), \doi{10.1103/PhysRevB.95.064405}.
\bibitem{note} Winding numbers label the homotopy classes of the torus. They measure the 2d flux through its two cycles.  
\bibitem{Newman} M.~E.~J.~Newman and G.~T.~Barkema, \textit{Monte-Carlo Methods in Statistical Physics}, Oxford 
  University Press, 1999.
\bibitem{notew} This is indeed verified for the sizes available, $L=1,\dots,8$. All winding numbers with $n^2 \leq 5$ are realized for $L=7,8$ (but not for $L<7$).
\bibitem{Rammal} R.~Rammal, G.~Toulouse, and M.~A.~Virasoro, Rev. Mod. Phys. \textbf{58}, 765 (1986), \doi{10.1103/RevModPhys.58.765}.
\bibitem{Nelson} D. Nelson, Phys. Rev. B \textbf{28}, 5515 (1983), \doi{10.1103/PhysRevB.28.5515}.
\end{thebibliography}
\end{document}